\documentclass[10pt, conference, letterpaper]{IEEEtran}
\IEEEoverridecommandlockouts
\usepackage{cite}
\usepackage{tikz, forest}
\usepackage{subcaption}[size=smaller]

\usepackage{amsmath,amssymb,amsfonts}
\usetikzlibrary{backgrounds}
\usetikzlibrary{matrix}
\usepackage{colortbl}
\usetikzlibrary{shapes.geometric}
\usetikzlibrary{shapes.arrows}
\usetikzlibrary{positioning}
\usetikzlibrary{shapes.geometric,backgrounds}
\usepackage{float}
\usepackage{enumitem}

\usepackage{url}
\usepackage{color,soul}

\usetikzlibrary{calc}
\usetikzlibrary{positioning}
\usetikzlibrary{decorations.text}
\usetikzlibrary{arrows.meta}
\usepackage{algorithmic}
\usepackage{graphicx}
\usepackage{textcomp}
\usepackage{xcolor}
\usepackage{multirow}

\def\BibTeX{{\rm B\kern-.05em{\sc i\kern-.025em b}\kern-.08em
    T\kern-.1667em\lower.7ex\hbox{E}\kern-.125emX}}
    
\begin{document}

    \title{\Large \bf SEER: Performance-Aware Leader Election in Single-Leader Consensus}

    \author{\IEEEauthorblockN{Ermin Sakic}
        \IEEEauthorblockA{
            \textit{Chair of Communication Networks} \\
            \textit{Technical University of Munich}\\
                Munich, Germany \\
                ermin.sakic@tum.de}
        \and
            \IEEEauthorblockN{Petra Vizarreta}
        \IEEEauthorblockA{
            \textit{Chair of Communication Networks} \\
            \textit{Technical University of Munich}\\
                Munich, Germany \\
                petra.vizarreta@tum.de}
        \and
            \IEEEauthorblockN{Wolfgang Kellerer}
        \IEEEauthorblockA{
            \textit{Chair of Communication Networks} \\
            \textit{Technical University of Munich}\\
                Munich, Germany \\
                wolfgang.kellerer@tum.de}
    }

\maketitle

\begin{abstract}

Modern stateful web services and distributed SDN controllers rely on log replication to omit data loss in case of fail-stop failures. In \emph{single-leader} execution, the leader replica is responsible for ordering log updates and the initiation of distributed commits, in order to guarantee log consistency. Network congestions, resource-heavy computation, and imbalanced resource allocations may, however, result in inappropriate leader selection and increased cluster response times. 

We present \texttt{SEER}, a logically centralized approach to performance prediction and efficient leader election in leader-based consensus systems. \texttt{SEER} autonomously identifies the replica that minimizes the average cluster response time, using prediction models trained dynamically at runtime. To balance the exploration and exploitation, \texttt{SEER} explores replicas' performance and updates their prediction models only after detecting significant system changes. We evaluate \texttt{SEER} in a traffic management scenario comprising $[3..7]$ Raft replicas, and well-known data-center and WAN topologies. Compared to the Raft's uniform leader election, \texttt{SEER} decreases the mean control plane response time by up to $\sim 32\%$. The benefit comes at the expense of the minimal adaptation of Raft election procedure and a slight increase in leader reconfiguration frequency, the latter being tunable with a guaranteed upper bound. No safety properties of Raft are invalidated by \texttt{SEER}.

\end{abstract}


\section{Introduction}
\label{introduction}

Critical services are often replicated and coupled into logical clusters, in order to tolerate fail-stop and/or Byzantine faults. Consensus protocols, such as \emph{leader-based} Raft~\cite{howard2015raft, ongaro2014search}, (Fast) Paxos~\cite{junqueira2007classic}, \emph{all-leader} Mencius~\cite{Mao:2008:MBE:1855741.1855767} and Clock-RSM~\cite{du2014clock}, ensure sequencing of client requests in an ordered log and its mirroring to replicas of the logical cluster. Incoming client requests are then executed in the order of their acceptance in the distributed log. Higher-layer applications can hence read and modify the replicated state records without awareness of the underlying consensus implementation. 

Leader-based Raft~\cite{howard2015raft, ongaro2014search} is often the consensus protocol of choice in production systems, e.g., in practical Kubernetes~\cite{netto2017state} / etcd~\cite{pedersen2018analysis}, Docker Swarm~\cite{Vohra2017} and Software Defined Networking (SDN) controller deployments, in particular in OpenDaylight and ONOS platforms~\cite{odl, berde2014onos}. Raft ensures strong consistent data replication, leader election and state reconciliation after failures. Prior to committing a log update, the leader in Raft serializes and proposes the update to remaining cluster replicas. After the majority of Raft cluster members have confirmed the update, it is \emph{committed} by the leader. Raft assumes the existence of \emph{terms}, where for each term, a single leader is elected by the majority of replicas. 

In its original specification, Raft relies on uniform leader election, where each replica is equally probable to become the candidate and eventually propose its leadership for the upcoming term. The uniform leader election in Raft can, however, lead to slow or inefficiently placed leaders and hence cause deceleration of the synchronization procedure and incur response time penalties for client requests. The issue of inefficient leader is exacerbated in geo-separated Raft deployments where network delays incurred by communication to the leader can cause high response times~\cite{xu2019elastic, kim2019load, liu2016leader, turcu2014general}. Similarly, recent IEC/IEEE 802.1 activities have motivated the support for dynamic network extensions in machine assemblies~\cite{indprofile, usecases}. These impose the requirement of Layer2 multicast tree establishment with response time guarantees using a reliable network controller (i.e., a Raft-enabled distributed SDN controller~\cite{odl, berde2014onos}). Thus, the response time of the distributed network control plane must be minimized to guarantee the upper bound value for tree computation and enforcement of $500ms$~\cite{usecases} in those scenarios.

Dynamic network incidents such as network failures or spanning tree reconfigurations~\cite{sakicresponse}, as well as high control plane load~\cite{hanmer2018friend, hanmer2019death} may result in leader or network overload on the path to the leader and thus deteriorate the resulting cluster performance. The few existing works that propse efficient leader-election either (i) focus on load balancing of controller replicas and rely e.g., on round-based leader election with minimal latency benefits~\cite{veronese2009spin}, (ii) consider the distributed commit delay as a function of network latency only~\cite{liu2016leader, eischer2018latency} or (iii) attempt to redesign Raft to alleviate the leader bottleneck~\cite{hanmer2019death, arora2017leader, xu2019elastic}. 

In this work, we instead make a case for model-based predictions of the best-performing leader (further \emph{best-leader}) at runtime and thus attempt to \emph{solve the problem of inefficient leader election}. To this end, we implement and evaluate a minimal adaptation of Raft leader election procedure, allowing for a remotely initiated candidate role assignment to arbitrary replicas. We thus remain compatible with the remainder of Raft protocol specification / reference implementations.

A logically centralized \texttt{SEER Elector} component periodically predicts the best-leader for the upcoming Raft term based on a dynamically trained per-replica performance prediction model. The prediction model is trained using the set of performance observations, related to resource utilization and observed round-trip and commit delays among replicas. The metrics are collected from dedicated end-point agents based on local knowledge only. Thus, we adopt a black-box approach to metrics collection, without relying on knowledge of detailed network parameters, in order to generalize \texttt{SEER}'s applicability to support arbitrary replicated services beyond SDN control plane. We minimize the number of reconfigurations necessary to explore and update the performance models of non-leader Raft replicas at runtime, by means of an unsupervised system novelty detection procedure that triggers performance exploration. 


\subsection{Our Contribution}
\label{contribution}

Succinctly, our contributions can be summarized as follows:
\begin{itemize}
	\item We motivate the leader election problem by a practical demonstration of the impact of resource contention, network dynamics, and heterogeneous resource allocation to nodes hosting the Raft replicas.
	\item We propose the usage of an online prediction model for enforcement of the best-leader. To this end, we evaluate three state-of-the-art regression methods and introduce the corresponding scoring metric.
	\item To tackle the trade-off between exploitation of predicted best-leader and exploration of non-leader's performance, we propose an unsupervised system state novelty detection method based on Local Outlier Factor (LOF)~\cite{Breunig:2000:LID:342009.335388}. Fitting of predictor models is executed only on detection of significant resource / network events. 
	\item We empirically evaluate and confirm the significant advantage of using the proposed prediction-based leader election compared to the uniform election procedure proposed by Raft's specification~\cite{howard2015raft, ongaro2014search}, in emulated Internet2 and Fat-Tree topologies, with varying replica cluster sizes and system parameterizations. We additionally quantify the \emph{tunable} increase in observed reconfiguration frequency, and with it, the imposed increase in cluster unavailability. Finally, we confirm the advantages and applicability of our approach in scenarios with imbalanced and balanced client request arrivals.
\end{itemize}

The rest of the paper is organized as follows: Sec. \ref{motivation} summarizes Raft and discusses the issues related to inefficient leader selection. Sec. \ref{modelanddesign} presents the system model and \texttt{SEER}'s design. Sec. \ref{eval} discusses the evaluated parameterizations and scenarios. Sec. \ref{results} presents the empirical results. \mbox{Sec. \ref{relatedwork}} discusses the related work. Sec. \ref{conclusion} concludes the paper.

\newcommand{\inputTikZ}[2]{%
	\scalebox{#1}{\input{#2}}  
}

\begin{figure*}
	\centering
	\scalebox{3.5}{\begin{tikzpicture}

	\coordinate (a) at (0,0);
	\coordinate (b) at (0,1.1);

	\coordinate (c) at (1,0);
	\coordinate (d) at (1,1.1);

	\coordinate (e) at (2.2,0);
	\coordinate (f) at (2.2,1.1);

	\coordinate (g) at (3.5,0);
	\coordinate (h) at (3.5,1.1);
	\coordinate (i) at (3.6,0);
	\coordinate (j) at (3.6,1.1);
	\coordinate (k) at (3.7,0);
	\coordinate (l) at (3.7,1.1);
	\draw (0, -0.2) -- (b)node[pos=1.1,scale=0.25]{Client} (1, -0.2) -- (d)node[pos=1.1,scale=0.25]{Source Replica (Follower)} (2.2, -0.2) -- (f)node[pos=1.1,scale=0.25]{Replica (Leader)} (3.5, -0.2) -- (h)node[pos=1.1,scale=0.25]{} (3.6, -0.2) -- (j)node[pos=1.1,scale=0.25]{Other Replicas (Followers)}(3.7, -0.2) -- (l)node[pos=1.1,scale=0.25]{};
	
	\draw[-{Stealth[scale=0.3]}] ($(a)!0.75!(b)$) -- node[above,scale=0.25,midway]{\footnotesize client\_request(\texttt{cmd})}($(c)!0.75!(d)$);

	\draw[-{Stealth[scale=0.3]}] ($(c)!0.74!(d)$) -- node[above,scale=0.25,midway]{\footnotesize proxy\_request(\texttt{cmd})}($(e)!0.74!(f)$);
	\draw[-{Stealth[scale=0.3]}] (2.2, 0.81) -- (2.3,0.81) -- node[right,scale=0.25,midway]{\footnotesize append\_log(\texttt{cmd})}(2.3,0.76) -- (2.2,0.76);

	\draw[-{Stealth[scale=0.3]}] (1, -0.02) -- (1.1, -0.02) -- node[right,scale=0.25,midway]{\footnotesize \texttt{commit\_idx++} \& exec(\texttt{cmd})}(1.1, -0.07) -- (1, -0.07);
	\draw[{Stealth[scale=0.3]}-] ($(c)!0.55!(d)$) -- node[above,scale=0.25,midway]{\footnotesize append\_entries(\texttt{cmd}, \texttt{prev\_log\_idx})}($(e)!0.6!(f)$);
	
	\draw[{Stealth[scale=0.3]}-] ($(g)!0.59!(h)$) -- node[above,scale=0.25,midway]{}($(e)!0.6!(f)$);
	\draw[{Stealth[scale=0.3]}-] ($(i)!0.55!(j)$) -- node[above,scale=0.25,midway]{\footnotesize append\_entries(\texttt{cmd}, \texttt{prev\_log\_idx})}($(e)!0.6!(f)$);
	\draw[{Stealth[scale=0.3]}-] ($(k)!0.51!(l)$) -- node[above,scale=0.25,midway]{}($(e)!0.6!(f)$);

	\draw[-{Stealth[scale=0.3]}] (1, 0.55) -- (1.1,0.55) -- node[right,scale=0.25,midway, align=center]{\footnotesize append\_log(\texttt{cmd}), \texttt{prev\_log\_idx++}}(1.1,0.5) -- (1,0.5);
	\draw[-{Stealth[scale=0.3]}] (3.5, 0.51) -- (3.8,0.51) -- node[right,scale=0.25,midway, align=center]{\footnotesize append\_log(\texttt{cmd}), \\ \footnotesize \texttt{prev\_log\_idx++}}(3.8,0.47) -- (3.5,0.47);
	\draw[-{Stealth[scale=0.3]}] (3.5, 0.51) -- (3.8,0.51) -- node[right,scale=0.25,midway]{}(3.8,0.47) -- (3.6,0.47);
	\draw[-{Stealth[scale=0.3]}] (3.5, 0.51) -- (3.8,0.51) -- node[right,scale=0.25,midway]{}(3.8,0.47) -- (3.7,0.47);

	\draw[-{Stealth[scale=0.3]}] ($(c)!0.35!(d)$) -- node[above,scale=0.25,midway]{\footnotesize \texttt{prev\_log\_idx}}($(e)!0.3!(f)$);
	\draw[-{Stealth[scale=0.3]}] ($(g)!0.31!(h)$) -- node[above,scale=0.25,midway]{}($(e)!0.3!(f)$);
	\draw[-{Stealth[scale=0.3]}] ($(i)!0.33!(j)$) -- node[above,scale=0.25,midway]{\footnotesize \texttt{prev\_log\_idx}}($(e)!0.3!(f)$);
	\draw[-{Stealth[scale=0.3]}] ($(k)!0.35!(l)$) -- node[above,scale=0.25,midway]{}($(e)!0.3!(f)$);

	\draw[-{Stealth[scale=0.3]}] (2.2, 0.25) -- (2.3,0.25) -- node[right,scale=0.25,midway]{\footnotesize \texttt{commit\_idx}++ \& exec(\texttt{cmd})}(2.3,0.2) -- (2.2,0.2);

	\draw[{Stealth[scale=0.3]}-] ($(c)!0.09!(d)$) -- node[above,scale=0.25,midway]{\footnotesize append\_entries(\texttt{commit\_idx})}($(e)!0.1!(f)$);
	\draw[{Stealth[scale=0.3]}-] ($(g)!0.09!(h)$) -- node[above,scale=0.25,midway]{}($(e)!0.1!(f)$);
	\draw[{Stealth[scale=0.3]}-] ($(i)!0.05!(j)$) -- node[above,scale=0.25,midway]{\footnotesize append\_entries(\texttt{commit\_idx})}($(e)!0.1!(f)$);
	\draw[-{Stealth[scale=0.3]}] (3.5, -0.02) -- (3.8,-0.02) -- node[right,scale=0.25,midway, align=center]{\footnotesize \texttt{commit\_idx}++ \& \\ \footnotesize exec(\texttt{cmd})}(3.8,-0.07) -- (3.5,-0.07);
	\draw[-{Stealth[scale=0.3]}] (3.5, -0.02) -- (3.8,-0.02) -- node[right,scale=0.25,midway]{}(3.8,-0.07) -- (3.6,-0.07);
	\draw[-{Stealth[scale=0.3]}] (3.5, -0.02) -- (3.8,-0.02) -- node[right,scale=0.25,midway]{}(3.8,-0.07) -- (3.7,-0.07);

	\draw[{Stealth[scale=0.3]}-] ($(k)!0.01!(l)$) -- node[above,scale=0.25,midway]{}($(e)!0.1!(f)$);

	\draw[{Stealth[scale=0.3]}-] ($(a)!-0.1!(b)$) -- node[above,scale=0.25,midway]{\footnotesize client\_response(\texttt{cmd})} ($(c)!-0.1!(d)$);

\end{tikzpicture}}  

	\caption{Sequence diagram of a successfully executed Raft log update commit. Assuming stable leadership, committing a log entry requires a single round trip to half of the cluster. Confirmations require $1.5$ round-trips on average, excluding the request transmission delays between the client and source replica.}
	\label{fig:raft_algorithm}
\end{figure*}
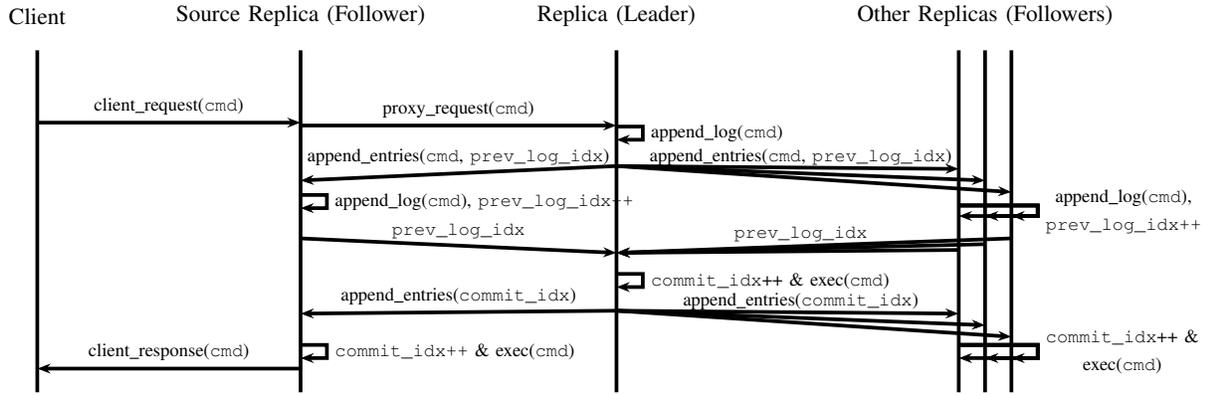

\section{Background and Motivation}
\label{motivation}
Raft is a distributed consensus algorithm that provides safe and ordered updates in a system comprised of multiple running replicas. It tries to solve the issues of understandability of the previous de-facto standard consensus algorithm Multi-Paxos~\cite{lamport1998part}, and additionally standardizes an implementation of leader election and post-failure replica recovery operations. In the following, we summarize selected design aspects of Raft.

\emph{Autonomous leader election~\cite{howard2015raft, ongaro2014search}}: The replicas in Raft may acquire the \emph{leader}, \emph{follower} or \emph{candidate} role. Raft distinguishes \emph{terms}, i.e., arbitrary time periods during which a unique leader is responsible for distributing client update requests to followers and ensuring safe commits. In the case of a leader failure, after an expiration of an internal follower timeout, the remaining followers automatically switch to a candidate role. A candidate is an active replica which offers to become the new cluster leader in the new Raft term. To do so, it propagates its candidate status to the other available replicas. If the cluster majority votes for the same candidate, the elected replica becomes the leader for the new term. 

\emph{Committing log updates in Raft}: Fig. \ref{fig:raft_algorithm} depicts a successfully executed log commit procedure. After receiving a command request, the source replica propagates the command to the current term leader. The leader proposes the according replicated log update to its followers, and additionally transmits the current log index (\texttt{prev\_log\_idx}). The followers with an up-to-date log respond with an incremented log index (\texttt{prev\_log\_idx++}), thus confirming the appended local log. After collecting the confirmations from at least half the active followers, the leader \emph{commits} the update to its commit log and increments its commit index (\texttt{commit\_idx++}). The leader then notifies the followers of the commit index update, which in return mark the update as committed and finally increment their own \texttt{commit\_idx} variable. Depending on the deployed application model, following a commit update, all replicas may execute the command and thus update their state machine (i.e., as replicated state machine instances). Alternatively, only a single replica or a subset of replicas execute the related command. 

For brevity, we omit the discussion of detailed structural and behavioral models of Raft in zero- and multiple-failure cases here, and instead refer the interested reader to~\cite{sakicresponse}. 

\begin{figure*}[htb]
        \centering
        \begin{subfigure}{0.33\textwidth}
                \begin{center}
                        \includegraphics[width=\textwidth]{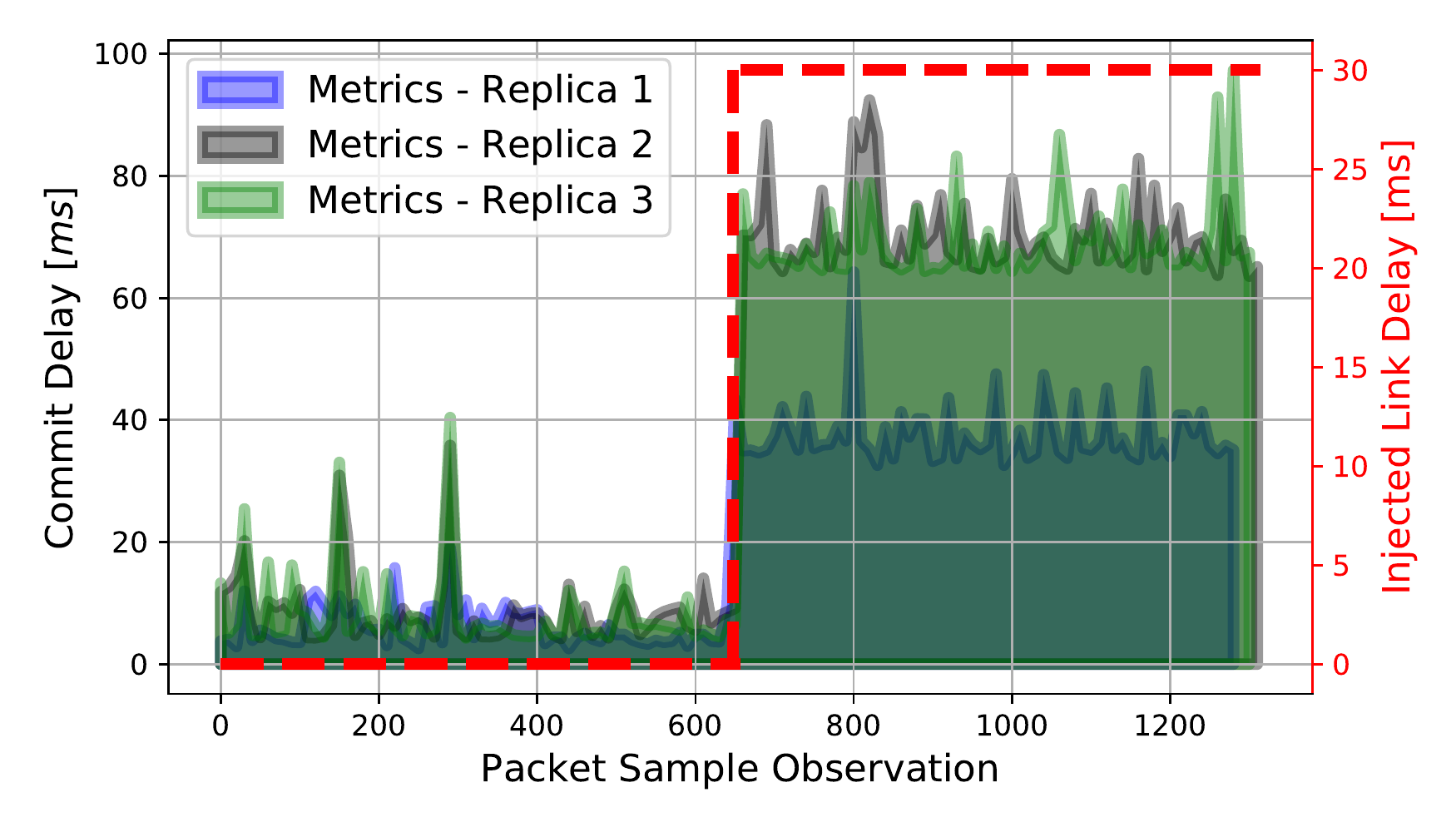}
                \end{center}
                \caption{Network Delay Injection}
                \label{fig:delayinject}
        \end{subfigure}%
        \begin{subfigure}{0.33\textwidth}
                \begin{center}
                        \includegraphics[width=\textwidth]{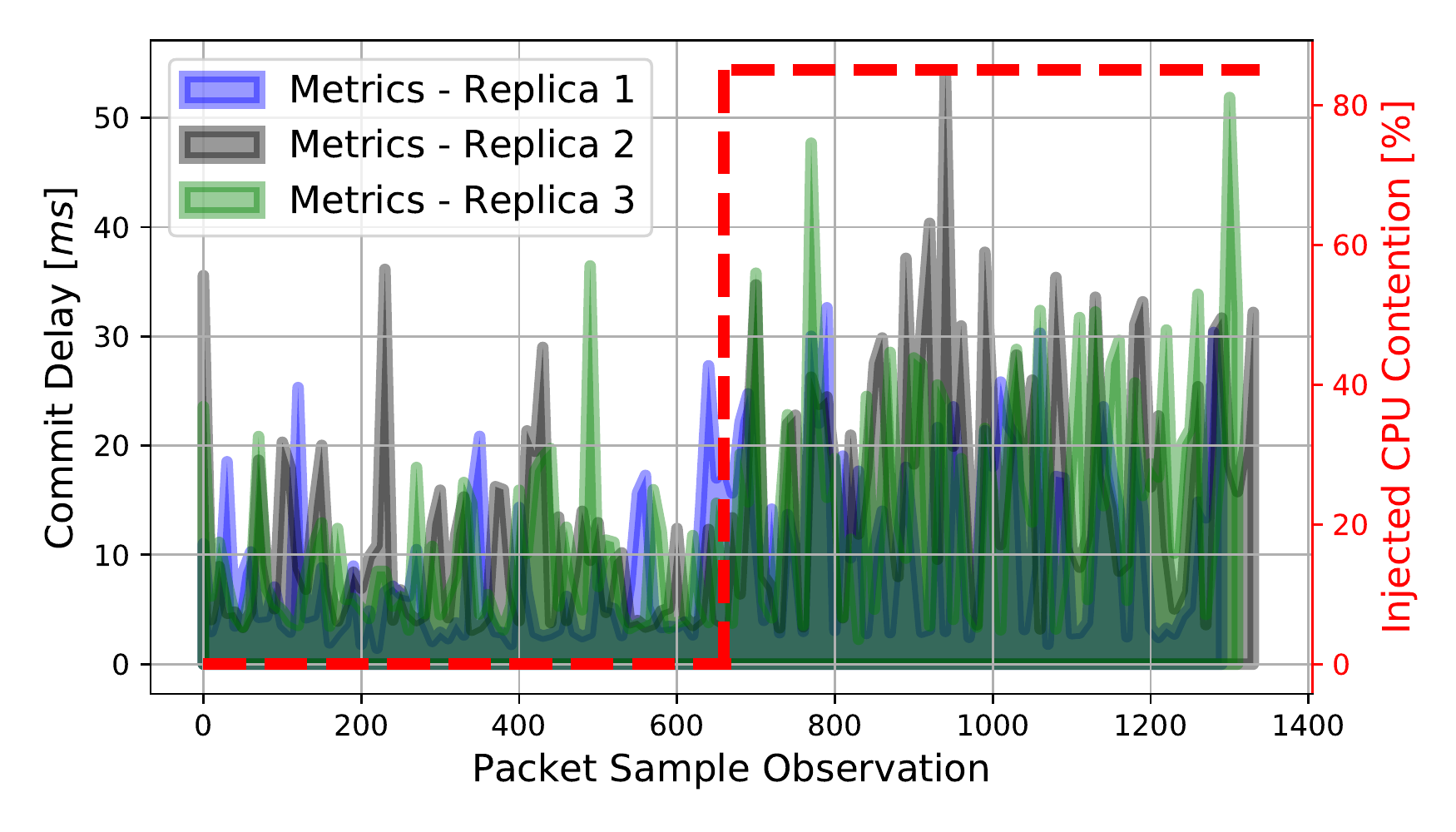}
                \end{center}
                \caption{Heavy CPU Workload Injection}
                \label{fig:contention_inject}
        \end{subfigure}%
        \begin{subfigure}{0.33\textwidth}
                \begin{center}
                        \includegraphics[width=\textwidth]{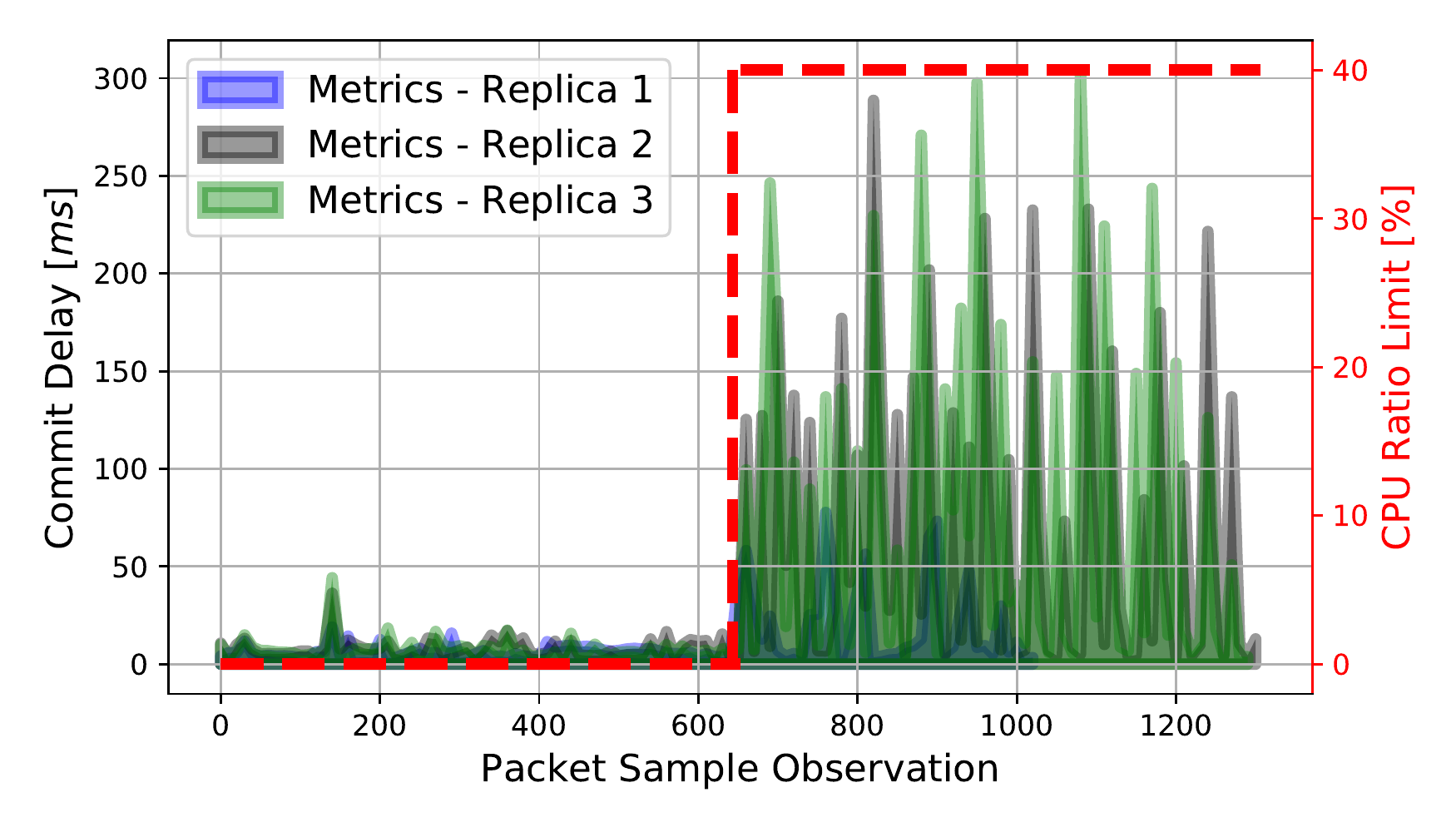}
                \end{center}
                \caption{Heterogeneous CPU Resource Allocation}
                \label{fig:limit_inject}
        \end{subfigure}
        \caption{Impact of sub-optimal leader election in a $3$-replica cluster. Depicted are the observed commit delays for distributed commits for randomly-placed replicas in a Fat-Tree topology with radix $k=4$ and $n=3$ levels~\cite{singla2016fat}. Events such as: (a) network contention and delays on the path to Raft leader; (b) CPU-heavy workloads; and (c) heterogeneous CPU resource allocation in the Raft leader; significantly deteriorate the overall cluster performance, thus motivating the need for adaptive best-leader election. In the depicted demonstrations, Replica 1 is the elected Raft leader (hence having the lowest observed commit delays), as well as the target of the contention events resulting in cluster performance deterioration.}
        \label{fig:metrics_impact}
\end{figure*}

\emph{Impact of sub-optimal leader election}: Fig. \ref{fig:metrics_impact} (a) demonstrates the negative impact of network contentions on the cluster performance - i.e., the observed per-transaction commit delay. System-under-test is a $3$-replica cluster, each executing an instance of  \texttt{PySyncObj}'s~\footnote{\texttt{PySyncObj} - A Python library for building fault-tolerant distributed systems - \url{https://github.com/bakwc/PySyncObj}} Raft implementation. Indeed, out-of-ordinary asymmetric network delay in the network attachment port of the leader replica deteriorates the read/write performance of the whole cluster. Fig. \ref{fig:metrics_impact} (b) demonstrates a similar impact after injecting additional CPU load in the leader replica's node, where 85\% of available CPU cycles are used by a contending process. Finally, Fig. \ref{fig:metrics_impact} (c) showcases the impact of heterogeneous resource mapping of the replicas. Limiting the available CPU cycles to 40\% of the originally available capacity in leader replica results in longer commit delays at each cluster replica, thus making a case for prioritizing high-performing replicas as leaders.

\section{Solution Design}
\label{modelanddesign}

\subsection{Model Overview}

We consider a distributed system comprising $|\mathcal{R}|$ Raft replicas, collected in a single cluster and deployed for the purpose of fault-tolerant operation. A configuration of $|\mathcal{R}| = 2F+1$ replicas tolerates a maximum of $F$ replica failures prior to cluster unavailability. Replicas agree on the order of updates and eventually commit the update requests in the per application replicated log, thus providing for strong consistency property (i.e., serializability and linearizability~\cite{linear}). We assume a non-Byzantine~\cite{eischer2018latency, sakic2018morph}, fail-stop~\cite{cachin2011introduction} model - replicas that fail, cease to work correctly.

\textbf{Network}: We assume Raft replicas connected in any-to-any manner with reliability provided using disjoint paths enabling for fail-over of replica-to-replica connections in case of network link/node outages. Alternatively, on link/node failures, redundant paths are reactively configured or alternative spanning trees are activated by e.g., re-converging (R)STP.  We assume arbitrary asymmetric delays between replicas, but also an eventual synchrony, in order to circumvent the FLP impossibility~\cite{flp}. 

\textbf{End-Points}: We assume loosely synchronized clocks among replicas (e.g., using NTP~\cite{mills1991internet}). Due to complexity of measuring unicast delay transmissions, we consider each replica to periodically transmit ICMP Echo requests to remote replicas and maintain the round trip time (RTT) values per replica pair. The clients transmit their requests to the closest/arbitrary replicas, i.e., \emph{source replicas}, which in return note the time instance of request arrival in the \emph{originating timestamp}, and forward the request to the leader of the current term for a subsequent majority confirmation. Following a confirmation of a committed request $q$, the source replica subtracts the \emph{originating} from \emph{current} timestamp and stores the difference as the \emph{commit delay} tuple $(q, d_q)$.  The replicas periodically transmit the observed RTTs, CPU/memory utilization metrics and the commit delay tuple to the logically centralized \texttt{SEER Elector} component.  We assume that each replica is capable of discovering the \texttt{SEER Elector}'s IP and port number.

\textbf{\texttt{SEER Elector}}: The logically centralized election function, responsible for: 
\begin{enumerate} 
	\item detection of system novelties and exploration of non-leader replica's performance; 
	\item subsequent collection of observation vectors from each Raft replica and refitting of their individual performance prediction models; 
	\item periodic prediction of the best-leader; and 
	\item enforcement of the best-leader. 
\end{enumerate} 

We assume reachability between \texttt{SEER Elector} and each other Raft replica. In case of the failure of the \texttt{SEER Elector} instance, a backup instance may take over its management role. In our implementation, \texttt{SEER} instance is implemented as a replicated process with leader election provided by a separate Raft session. \texttt{SEER Elector} is hence bootstrapped without additional user interaction. Failure of a \texttt{SEER Elector} instance results in no intermediate unavailability of the monitored / managed Raft sessions.

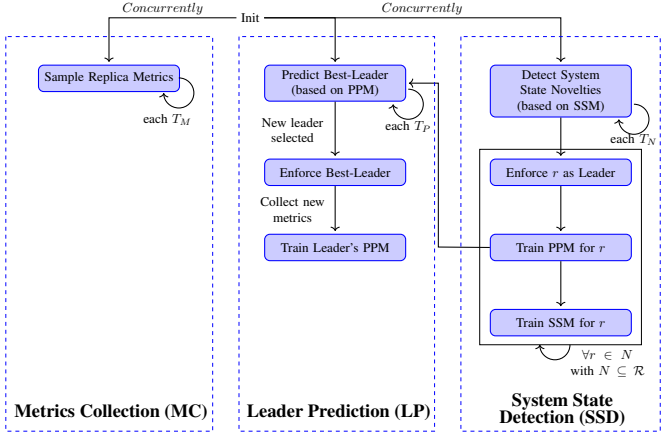
\begin{figure}[htb]
	\centering
	\scalebox{0.5}{\tikzset{
  mybackground/.style={execute at end picture={
        \begin{scope}[on background layer]
          \draw[black!15,fill=black!5,rounded corners=1ex] (current bounding box.south west)
                    rectangle (current bounding box.north east);
        \end{scope}
    }},
}

\begin{tikzpicture} [
    decision/.style = { diamond, draw=blue, thick, fill=blue!20,
                        text width=5em, text badly centered,
                        inner sep=1pt, rounded corners },
    block/.style    = { rectangle, draw=blue, thick, 
                        fill=blue!20, text width=10em, text centered,
                        rounded corners, minimum height=2em },
    line/.style     = { draw, thick, ->, shorten >=2pt }, node distance = 3cm, auto
  ]

    \node [label] (start) {Init};
    \node [block=above:sample] (sample) [below left=1cm and 1.5cm of start] {Sample Replica Metrics};
    \node [block=right:sample] (predict)  [below right=1cm and 0cm of start] {Predict Best-Leader (based on PPM)};
    \node [block=right:predict] (nd) [below right=1cm and 6cm of start] {Detect System State Novelties (based on SSM)};
    \node [block=below:predict] (expert) [below right=3.5cm and 0cm of start] {Enforce Best-Leader};
    \node [block=below:expert] (update2) [below right=5.5cm and 0cm of start] {Train Leader's PPM};
    \node [block=below:nd] (update)  [below right=3.5cm and 6cm of start] {Enforce $r$ as Leader};
    \node [block=below:nd] (update3)  [below right=5.5cm and 6cm of start] {Train PPM for $r$};
    \node [block=below:update3] (update4) [below right=7.5cm and 6cm of start] {Train SSM for $r$};
    \node [label, left of=update2] (test) {};
    
    \path [line] (update3) --  (update4);
    
    \path [line] (predict) --  node [text width=2cm, midway, left, align=center] {New leader selected} (expert);
    \path [line] (nd) -- (update);
    \path [line] (expert) -- node [text width=2cm, midway, left, align=center] {Collect new metrics} (update2);
    \path [line] (update) -- (update3);
    \path [line] (update3.west) -- (5,-6.1) |- (predict.east);
	\path [line] (start) -| (sample.north);
    \path [line] (start) -|(predict.north);
    \path [line] (start) -| node [text width=1cm, midway, above, align=center] {} (nd.north);
    \node [label, text width=1cm, align=center] at (-2.9, 0.25) {$Concurrently$};
    \node [label, text width=1cm, align=center] at (4, 0.25) {$Concurrently$};
    \draw[black,thick,->] (-1.9,-1.6) arc (90:-180:4mm);
    \draw[black,thick,->] (4.2,-1.9) arc (90:-180:4mm);
    \draw[black,thick,->] (10.2,-2.3) arc (90:-180:4mm);
    
    \draw[black,thick,->] (8.5,-8.7) arc (0:-180:4mm);
    \node[label, text width=3cm, align=center] at (9.5, -9.2) {$\forall r \in N$\\ with $N \subseteq \mathcal{R}$};   
    \draw [draw=black] (6.1,-3.5) rectangle (10.4,-8.7);

    \draw [draw=blue, dashed] (-6.5,-0.5) rectangle (-1, -11);
    \draw [draw=blue, dashed] (-0.3,-0.5) rectangle (4.9, -11);
    \draw [draw=blue, dashed] (5.6,-0.5) rectangle (10.9, -11);
    \node[label] at (-3.7, -10.5) {\Large \textbf{Metrics Collection (MC)}}; 
    \node[label] at (2.35, -10.5) {\Large \textbf{Leader Prediction (LP)}}; 
    \node[label={[align=center]\Large \textbf{System State} \\ \Large \textbf{Detection (SSD)}}] at (8.3, -11.1) {}; 
        
    \node[label] at (-2.2, -2.7) {each $T_M$}; 
    \node[label] at (4.2, -2.9) {each $T_P$}; 
    \node[label] at (10.2, -3.25) {each $T_N$}; 
\end{tikzpicture}}  

	\caption{LP predicts the best-leader based on the latest collected samples for each candidate and periodically elects the best-leader. SSD uses all collected sample observations during the previous sampling period and periodically refits the per-replica prediction models (PPMs). Additionally, on detection of system state novelties, SSD explores each candidate node for which a configuration novelty was detected (positive triggers for majority of observations) and similarly retrains the PPMs. It additionally adjusts the System State Model (SSM) according to the newly observed system state.}
	\label{fig:overall_algorithm}
\end{figure}

Fig. \ref{fig:overall_algorithm} gives the overview of concurrent interactions between the Metrics Collection (MC), the Leader Prediction (LP) and System State Detection (SSD) processes of \texttt{SEER Elector}. Note that LP and SSD blocks execute exclusively in \texttt{SEER}, while MC has a counterpart agent in each Raft replica of the system dedicated to exposing the reporting samples. We next discuss the proposed blocks in more detail.

\subsection{Prediction of Best-Leader Replica}

\renewcommand{\vec}[1]{\mathbf{#1}}

\textbf{Metrics Collection (MC)}: Each $T_M$ time instants replicas transmit their current observation vectors to \texttt{SEER Elector}. An observation vector $\vec{o}^{r_i} = [o^{cpu}_{r_i}, o^{ram}_{r_i}, o^{rtt}(r_i,r_0), \hdots,  o^{rtt} (r_i,r_{|\mathcal{R}|})]^\textbf{T}$ comprises the simple moving average (SMAs) of replica $r_i$'s measured resource utilization metrics (i.e., in our case, occupied CPU / RAM resources) and RTT values to remaining replicas in set $\mathcal{R} \setminus r_i := \{x \in \mathcal{R} | x \neq r_i\}$. We chose SMAs in order to reduce the noise of each report. Additionally, the current term leader $l$ asynchronously transmits  messages $\vec{o}^l_q = (\vec{o}^l, q)$ to \texttt{SEER Elector} in bulk or whenever a new distributed commit $q$ is processed by $l$. For each successfully processed commit $q$, the source replica $src(q)$ transmits the measured commit delay $d_q^{src(q)}$ including its globally unique request identifier to \texttt{SEER Elector}.

\textbf{Leader Prediction (LP)}: Each $T_P$, \texttt{SEER Elector} feeds the most recent observation vector $\vec{o}^r$ from each replica $r$ into its declared predictor, and a per-replica prediction model (PPM) is used to predict the resulting commit delay performance of replica candidate $r$ as potential leader for the upcoming term $t$. The sum of predicted commit delays for replica $r$ as potential leader of term $t$ denotes its fitness score
\begin{equation*}
f(r,t): \sum_{\forall k \in \mathcal{R}} \frac{\bar{y_k^r}}{|\mathcal{R}|}.
\end{equation*}
Here, $\bar{y_k^r}$ denotes the predicted delay for a request initiated at source-replica $k$, assuming $r$ takes over the leadership. 

\texttt{SEER}'s election procedure elects the leader replica $R_{min}^t$ set, for target term $t$ that contains the set of best-leaders that minimize the predicted score, i.e.,
\begin{equation*}
R_{min}^t = \{r | f(r, t) = \min_{r \in \mathcal{R}} f(r,t)\}.
\end{equation*}

Term duration (i.e., the leader re-election period) is lower-bounded by $min(T_P, T_N)$, where $T_N$ represents the period of system state novelty detection (discussed in Sec. \ref{exploration}). We note that $T_M$ and $T_N$ may be tuned to prevent too frequent / seldom leader adaptation. We next discuss the PPM training procedure and introduce exemplary predictors.

\subsection{Training of Per-Replica Performance Prediction Models}

After collecting commit delay reports  $d_i^{src(i)} \hdots d_z^{src(z)}$ from all replicas (at least one commit report from each replica), \texttt{SEER Elector} computes the means of individual resource-related features of leader's observation vectors $\vec{o}^l_i\hdots\vec{o}^l_z$ that match the commit request identifiers $i .. z$. It additionally extracts the RTTs $o^{rtt}(l, src(i)) \hdots o^{rtt}(l, src(z))$ observed between leader $l$ and source replicas at the time of request processing. The mapping  of pre-processed leader $l$'s observation input vector and the commit delay vector then constitutes a single training sample:

\begin{small} 
\begin{align*}
\small
\vec{x} &= \begin{bmatrix}
\frac{1}{|\mathcal{R}|} \sum_{q = i}^z o_{l,q}^{CPU} \\
\frac{1}{|\mathcal{R}|} \sum_{q = i}^z o_{l,q}^{RAM} \\
o_{i}^{rtt} (l, src(i)) \\
\vdots \\
o_{z}^{rtt} (l, src(z)) 
\end{bmatrix}; 
\vec{y}=\begin{bmatrix}
d_{i}^{src(i)} \\
\vdots \\
d_{z}^{src(z)}
\end{bmatrix}; (l, src(i)\ ..\ src(z)) \in \mathcal{R},
\end{align*}
\end{small}

with $|\vec{x}| = |\mathcal{R}| + 2$ and $|\vec{y}| = |\mathcal{R}|$.

In the evaluation of the generic model applicability, we train and leverage the following exemplary predictors:

\begin{itemize}
\item Ordinary Least Squares (OLS): OLS identifies a linear function of the feature vector that minimizes the sum of squared differences between the observed dependent variable and the variable predicted by the linear function. Loss $\epsilon$ is characterized by mean squared error (MSE) as
	\begin{equation*}
		\epsilon(\vec{y}, \bar{\vec{y}}) = \frac{1}{n} \sum_{i=1}^{n}(y_i-\bar{y_i})^2  = \frac{1}{n} \sum_{i=1}^{n}(y_i-x_i\beta)^2,
	\end{equation*}
where $\beta$ denotes the regression coefficients.	

\item Elastic Net (EN)~\cite{zou2005regularization}: EN extends the OLS with LASSO (\texttt{L1})~\cite{lasso} and ridge (\texttt{L2})~\cite{ridge} regularization terms, to prevent over-fitting
	\begin{equation*}
		\epsilon_{\texttt{L1,L2}}(\vec{y}, \bar{\vec{y}})
		= \epsilon(y, \bar{y}) + \lambda((1-\alpha)\sum_{j=0}^{m}\beta_m^2 + \alpha\sum_{j=0}^{m} |\beta_m|).
	\end{equation*}

\item Artificial Neural Network (ANN): ANNs can be trained to produce predictions by training the weights on input data. Learning is performed via back-propagation that adjusts the weights and finds an optimum which minimizes the prediction error based on observed target output and predicted variable. Due to their lightweight implementation and high accuracy, in the past, ANNs were used in core performance prediction and thread partitioning strategies~\cite{nemirovsky2017machine, li2015using, shulga, shekhar}. 
\end{itemize}

Updates to ANN require adjusting the model weights while updates to OLS and EN models necessitate recomputation of regression coefficients. \texttt{SEER Elector} trains the PPMs dynamically at runtime: (i) current term leader's PPM are updated periodically each $T_P$; (ii) all replicas' PPMs are updated periodically each $T_N$ time instants whenever system state novelties have been detected (ref. Section \ref{exploration}); where $T_P \ll T_N$. The intermediate predictions thus result in new leader's election and training of its PPM more frequently than the full exploration procedure. In the following section, we discuss the exploration in more detail.

\subsection{Exploration of Followers' Performance}
\label{exploration}
Raft enforces the leadership of a single replica at any time. Hence, training of the predictor's PPM with ground truth commit delay values is efficiently possible for the current leader only. The remaining replicas must eventually be also elected as leaders in order to guarantee eventual updates of their PPMs. In case of rare explorations, followers' PPMs may become outdated. To provide for a good trade-off between exploration of replicas' performance and exploitation of the best-leader, we propose an unsupervised system change detection process that guarantees eventual exploration of all replicas.

\emph{Note}: Indeed, additional $|\mathcal{R}|-1$ parallel Raft sessions could be executed with each replica as leader in one of the sessions, in order to allow for parallel collection of ground-truth values and concurrent PPM adjustment. This would, however, impose a higher overhead of training and Raft session replications.

\emph{We explore the performance of replicas only when the observed system state of the replica has changed relative to the previous observation window}. Changes in resources (SMAs of CPU/RAM utilizations and RTTs) are considered the triggers for system updates. To model the system state we proceed as follows in the SSD block (ref. Fig. \ref{fig:overall_algorithm}).

\emph{Populating the initial System State Model (SSM)}: To discover system state novelties, we rely on unsupervised outlier detection using Local Outlier Factor (LOF)~\cite{Breunig:2000:LID:342009.335388}. Following the collection of network metrics during an ongoing term, \emph{per-replica} SSM models are generated. For each observation sample (comprising resource and RTT metrics) submitted by a replica, an anomaly score is computed, defining the isolation of the sample to the surrounding neighborhood (i.e., other input samples). LOF relies on $k$-nearest neighbors to determine locality of a sample, whose distance is used to estimate the local density of the sample. By comparing the local density of observed samples to those of its $k$-nearest neighbors, samples with lower density are identified and pruned in the filtering step. After removing the outliers, we store the remaining neighborhood samples in the replica-specific SSM model.

\textbf{System State Detection (SSD)}: Each $T_N$, outliers are filtered out of the next set of collected samples. The remaining non-pruned samples are evaluated against $k$ neighbors in the current SSM. If sufficient individual deviations are detected in the new sample set (e.g., $>98.5\%$ of observed samples in our evaluation), the latest samples constitute the new neighborhood and thus the new SSM. The new SSM hence defines the new active system state. In case a new SSM must be generated, that replica's performance as leader is explored, i.e., its PPM is retrained. To this end, \texttt{SEER} proceeds to elect the triggering instance as the leader of a temporary term, and evaluates its performance during a short metrics collection period (as depicted in SSD block of Fig. \ref{fig:overall_algorithm}).

\begin{figure}[htb]
	\centering
	\begin{center}
		\includegraphics[width=0.4\textwidth]{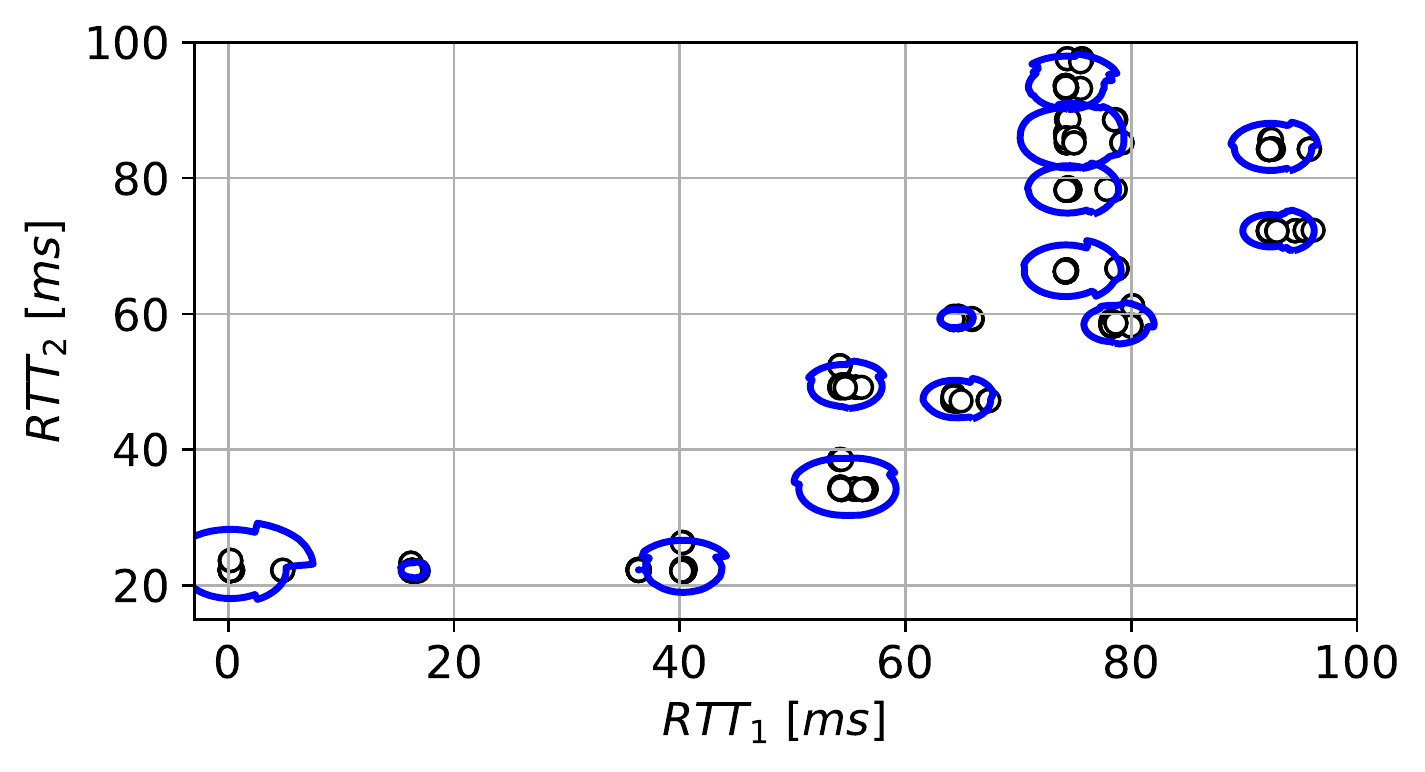}
	\end{center}
	\caption{Visual representation of system states observed by a replica after $15$ intentional delay injections in an emulated network deploying $3$ Raft replicas. For brevity, only the two RTT dimensions are depicted. Blue borders are the frontiers of the identified neighborhood uniquely describing the current SSM. }
	\label{fig:lof_rtts}
\end{figure}

Fig. \ref{fig:lof_rtts} depicts the simplest 2D scatter plot of detected neighborhoods in a 3-node replica cluster (each representing a unique system state). The exemplary system states are defined by the observing replica's RTT relation to the neighboring replicas only. Blue borders depict the frontier, delimiting the distribution of actual system state observations (RTTs). Samples which are inliers (matching the neighborhood) are the expected samples, while samples lying outside the borders of current active state are considered either novelty samples or outliers. Namely, future arriving samples, lying outside the current active frontier, are considered novelty samples if the majority of detected samples density does not deviate from their $k$-neighbors. These samples are thus taken into account in the definition of the subsequent SSM. The samples not mapped to either the current system state or neighborhood of the follow-up state are the outliers and are not depicted here.

In practice, determining the SSM requires evaluating and comparing multi-dimensional input vectors comprising additional resource features (i.e., CPU/RAM utilization) and RTT values to all other replicas as well. For simpler visualization, we do not depict more than two dimensions of the state model here. In implementation and evaluation, however, \texttt{SEER} considers the multivariate observation samples in SSM neighborhood definition and $k$-neighbor distance computation.

\subsection{Mitigation of Catastrophic Forgetting}

The proposed approach to predictor training aims to incrementally improve its predictions for the current leader's model after each leader election, and for all other replicas latest after each detected novelty. Training the PPMs based on recent observation samples alone, however, may load to catastrophic forgetting of previously learned relations. To mitigate this,  we mix the data from earlier sessions with the current session' samples ("rehearsal learning"~\cite{kemker2018measuring}). Thus, old samples are shuffled with the new samples so to appear independent in the training batches. Processing all old data during training would lead to monotonically increasing computation complexity, hence we bound the training complexity by randomly sampling a configurably-sized subset of old data and subsequently evaluate the long-term improvement to predictions.

\subsection{Upholding Raft's Safety Guarantees}

The leader of a new Raft term is elected only if it contains the committed entries from previous terms (ref. \S5.4.1 of~\cite{ongaro2014search}). Voters vote for a target leader candidate only if their committed entries are present on the target candidate. Thus, following the election of a new leader, there exists no need to transfer any previously committed entries to the leader.

\texttt{SEER} relies on Raft's voting process to prevent an inconsistent candidate from winning an election. It does not invalidate any of the safety guarantees due to how a leader is enforced: Best-leader candidate elected by \texttt{SEER} attempts collecting the majority vote for the following term, and in the case of a failed attempt, increments the term and indefinitely retries the voting step, until either elected or notified of cancellation by the \texttt{SEER Elector} (e.g., due to another replica being elected as the best-leader). The initial election attempts may hence fail due to inconsistency of the best-leader log compared to that of the leader of the previous active term. Eventually, however, the elected best-leader commits the outstanding updates to its log and becomes viable for election.  

\section{Evaluation Methodology}
\label{eval}

To evaluate the proposed model, we rely on and extend \texttt{PySyncObj}, a Python library for building distributed fault-tolerant replicated state machines. \texttt{PySyncObj} implements the reference Raft specification, and provides for straightforward tuning of Raft parameters (i.e., the candidate/follower timeout, log compaction frequency etc.). To evaluate the benefit of best-leader prediction, we extend \texttt{PySyncObj} with an interface for remote initiation of a replica's leadership and compare the prediction-based method with the Raft's native leader election relying on uniform timeout \cite{howard2015raft, ongaro2014search}.

\textbf{Test Application \& Clients}: To evaluate the proposed model, we implement a routing application and replicate it across $|\mathcal{R}| = [3..7]$ Raft replicas. After receiving a user request (command) comprising two randomly selected end-points, the current Raft leader distributes the update request to the commit log and, following a majority confirmation, computes the corresponding route using Dijkstra's SP algorithm. For simplicity, we assume that for each Raft replica, a single client is hosted on the same server node. Clients correspondingly select the closest replica as the \emph{source} replica.

The incoming client request arrivals follow a negative exponential distribution (n.e.d.) \cite{huang2017dynamic}. To reason about the prediction performance for imbalanced client workloads, we consider two distributions for the \emph{rate parameter} $\lambda$: (i) a \emph{balanced} equal rate for each client / source replica with $\lambda_{r_0} = \hdots = \lambda_{r_{|\mathcal{R}|}} = \lambda^B, \forall {r_0, ..., r_{|\mathcal{R}|}} \in \mathcal{R}$; and (ii) \emph{imbalanced} rates, where rates of individual clients correspond to uniformly distributed multiples of the base rate $\lambda^B$: $\lambda_{r_i} = \frac{1}{m_{r}} \lambda^B$. The imbalanced arrival rates expose the prediction performance when availability of client reports is limited.

 \newcolumntype{P}[1]{>{\centering\arraybackslash}p{#1}}
 \newcolumntype{M}[1]{>{\centering\arraybackslash}m{#1}}

 \definecolor{Gray}{gray}{0.9}
 \begin{table}[htb]
 	\scriptsize
 	\centering
 	\caption{Evaluation Parameters}
 	\resizebox{0.5\textwidth}{!}{%
 	\begin{tabular}{M{0.9cm}|M{3cm}|M{4.3cm}}
 		\textbf{Param} & \textbf{Value} & \textbf{Comment} \\
 		\hline
		\hline
 		$|\mathcal{R}|$ & $[3, 5, 7]$ & Raft cluster size \\
 		\rowcolor{Gray}
		N/A & {ANN, EN, OLS} & Predictor selection \\
 		$I$ & {1 (True), 0 (False)} & Imbalance in client requests \\
 		 		\rowcolor{Gray}
 		$P$ & {1 (True), 0 (False)} & Intermediate predictions \\
 		$H$ & {1 (True), 0 (False)} & Learning with rehearsals \\
 		 		\rowcolor{Gray}
 		$T_I$ & $5s$ & Period - Delay / CPU workload injections \\
 		$T_P$ & $2s$ & Period - LP block execution (ref. Fig. \ref{fig:overall_algorithm})  \\
 		 		\rowcolor{Gray}
 		$T_N$ & $20s$ & Period - SSD block execution (ref. Fig. \ref{fig:overall_algorithm})  \\
 		$J_{CPU}$ & 
 		$|\mathcal{R}| = 3: U(70, 90) \% \newline
 		|\mathcal{R}| = 5: U(65, 85) \% \newline
 		|\mathcal{R}| = 7: U(60, 80) \%$
 		& Injected per-replica CPU contention \\
 		 		\rowcolor{Gray}
		$J_{D}$ & $U(5, 35) ms$ & Injected uplink network delay \\
		$J_{LIM}$ & $|\mathcal{R}| = 3: U(10, 50) \% \newline
		|\mathcal{R}| = 5: U(15, 55) \% \newline
		|\mathcal{R}| = 7: U(20, 60) \%$
		& Per-replica CPU resource bound \\
				 		\rowcolor{Gray}
		$\frac{1}{\lambda^{B}}$ & $[50, 70] ms$ & Per-client request arrival rate \\
		$m_{r}$ & $U(2,5)$ & Rate multiplier for imbalanced requests \\
 		\hline
 	\end{tabular}
 }
\label{other_params}
 \end{table}

 \textbf{System state change triggers}: We distinguish three types of external triggers: (i) delay injections, i.e., uniformly distributed asymmetric delays in the attachment port of the target Raft replica; (ii) contending CPU-heavy workloads, executed on the same node as the target Raft replica; and (iii) heterogeneous resource allocations, that limit the maximum achievable performance of the target replica. 
 
 Intensities of delay, workload injections and resource allocations are uniformly distributed as per Table \ref{other_params}. The injections occur periodically each $T_I$. Each parameterization is evaluated over a course of $30$ state change injections. Each Raft replica is deployed in a Docker container, with a physical CPU core allocated exclusive for isolated execution (using \texttt{cpuset}). To enforce heterogeneous resource allocations, we rely on \texttt{cpulimit}'s upper bound allocation for the target core (i.e., target replica). To achieve a fair comparison of all predictors, we randomly generate the injection traces and replay them for different evaluated parameterizations (i.e., a unique injection / allocation trace per evaluated Raft cluster size).
 
\textbf{Network}: \texttt{SEER} was evaluated against an emulated network comprising a number of interconnected Open vSwitch v2.11.0 instances isolated in individual Docker containers.  The deployed network topologies connecting the distributed Raft replicas are either Internet2 ($34$ switches) or Fat-Tree with radix $4$ and $3$ levels ($20$ switches) \cite{singla2016fat}. To reflect the delays incurred by the length of the optical links in the geographically scattered Internet2, we assume a travel speed of light of $2 \cdot 10^6 \frac{km}{s}$ in the optical fiber links. We derive the link distances from the publicly available geographical Internet2 data\footnote{Internet2 topological data (provided by POCO project) - \url{https://github.com/lsinfo3/poco/tree/master/topologies}} and inject the corresponding propagation delays using Linux’s \texttt{tc} tool. In contrast, the links of the Fat-Tree only posses the inherit processing and queuing delays. In the Internet2, we leverage a Raft replica placement that allows for high robustness against replica failures according to \cite{sakic2018morph, hock2014poco}. In the case of Fat-Tree, replica / client pairs were placed on leaf-nodes in Round-Robin order similar to \cite{sakic2018morph}.

\textbf{PPM model computation:} If a feature has a variance that is orders of magnitude larger than others, it might dominate EN's objective function due to values with smaller amplitudes being penalized more by \texttt{L1} and \texttt{L2} regularizers, in result making the estimator unable to learn from all features. Hence, prior to fitting the models, we standardize the historical data's metrics to $\mathcal{N}(0,1)$. To account for non-linear relationship between the input features and the commit delay output, in case of EN we extend the feature vector with polynomial and interaction features. Thus, the transformed input vector used in training and prediction comprises all polynomial combinations of the features with a degree less than or equal to $2$. 

The ANN's predictor uses the well-known ADAM optimizer~\cite{kingma2014adam} on a neural network composed of $3$ hidden layers (ref. Table \ref{params}). We use exponential linear unit (ELU) as activation function of hidden units and linear output layer (predicted commit delay values are unbounded). Mean squared error (MSE) is used as the error minimization function. 

An overview of optimal hyperparameters detected by grid-search for both ANN and EN predictors is presented in \mbox{Table \ref{params}}. The optimal parameters were established by observing the minimized sum of MSEs of commit delay predictions after $20$ consecutive test runs per configuration.

\begin{table}[H]
	\centering
	\caption{Grid Search Evaluation of Predictor Parameters}
	\begin{tabular}{M{2cm}|M{2.7cm}|M{1.7cm}}
		\textbf{Predictor} & \textbf{Parameter} & \textbf{Value} \\
		\hline
		\hline
		\multirow{2}{*}{ANN}
		& No. hidden layers & $2,  \textbf{3}, 4$ \\
		& No. hidden units & $20, \textbf{50}, 100$ \\
		\hline
		\multirow{2}{*}{Elastic Net}
		& Intercept & \textbf{True}, False \\
		& $\alpha$ (\texttt{L1}-Term Ratio) & $0.1, \textbf{0.5}, 0.9$ \\
	\end{tabular}
	\label{params}
\end{table}

\textbf{SSM model computation:} To detect $k$-nearest neighbors, we rely on a neighborhood search algorithm using k-d trees~\cite{muja2014scalable} and the Local Outlier Factor (LOF \cite{Breunig:2000:LID:342009.335388}) implemented in \texttt{scikit-learn} \cite{pedregosa2011scikit}. The neighborhood size was set to $k = 20$ as per general recommendation in \cite{lofdetection}. 

\textbf{Computing resources}: We evaluate \texttt{SEER} on a recent $8$-core AMD CPU comprising one core dedicated to system processes, and up to $7$ cores used in isolated execution of Docker containers hosting the Raft replicas. Up to $20$ ($34$) Fat-Tree (Internet2) vSwitches uniformly utilize the unused CPU resources, but due to a low raw bandwidth utilization by \texttt{SEER} and Raft synchronization, their impact on CPU resources is negligible. \texttt{SEER} was fully implemented in Python. Predictors rely on \emph{TF 2.0}, \emph{Keras} and \emph{scikit-learn} for most operations.

\section{Results}
\label{results}

\subsection{Impact of System Dynamics}

\begin{figure*}[htb]
        \centering
        \begin{subfigure}[t]{0.08\textwidth}
        		\includegraphics[width=0.6\textwidth]{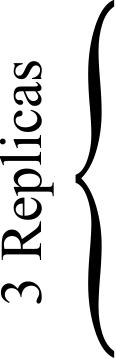}
        		\includegraphics[width=0.6\textwidth]{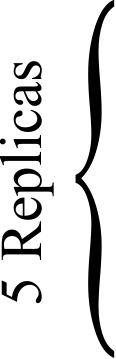}
        		\includegraphics[width=0.6\textwidth]{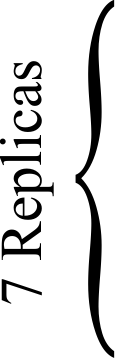}
        \end{subfigure}%
        \begin{subfigure}[t]{0.3\textwidth}
                \begin{center}
                        \includegraphics[width=\textwidth]{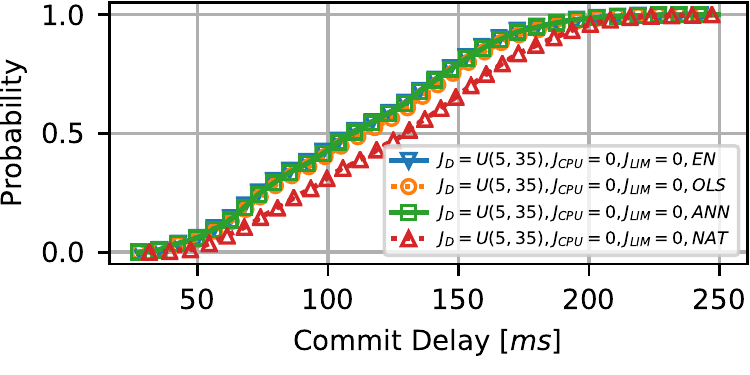}
                        \\
                        \includegraphics[width=\textwidth]{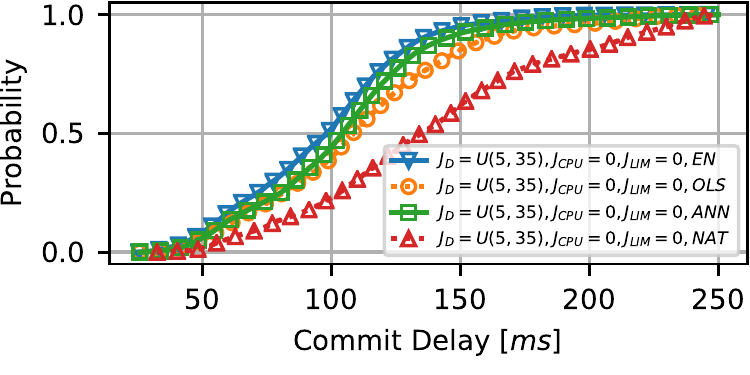}
                        \includegraphics[width=\textwidth]{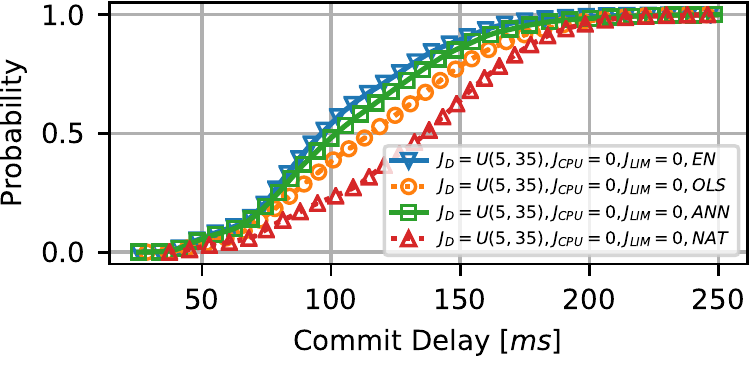}
                \end{center}
                \caption{Network Delay Injections}
                \label{fig:impact_delay}
        \end{subfigure}%
        \begin{subfigure}[t]{0.3\textwidth}
                \begin{center}
                        \includegraphics[width=\textwidth]{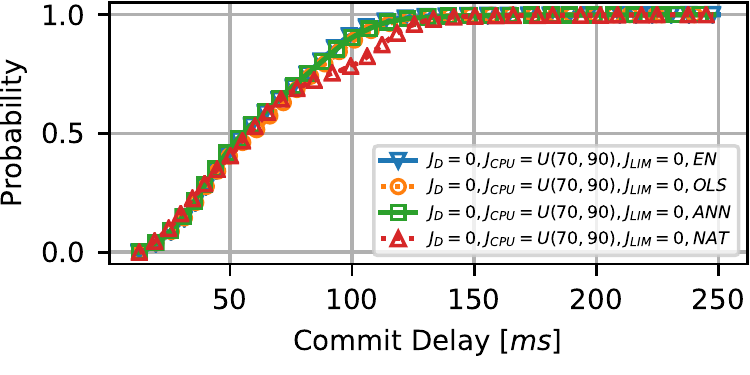}
                        \includegraphics[width=\textwidth]{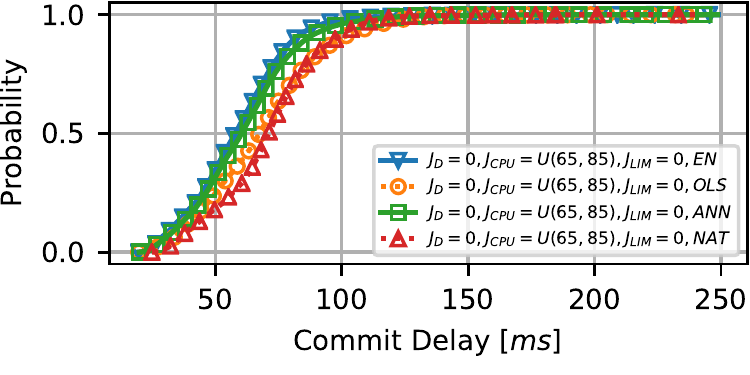}
                        \includegraphics[width=\textwidth]{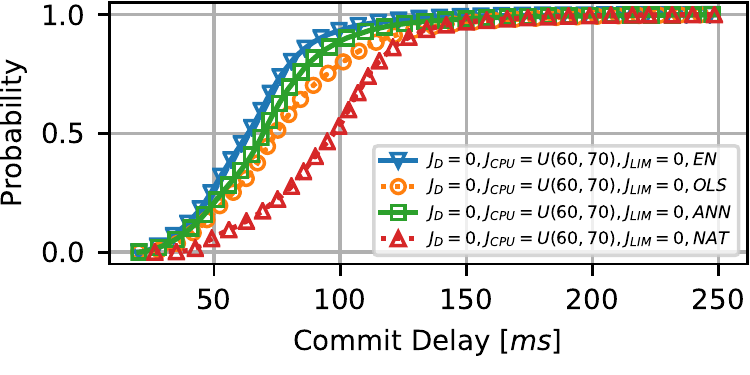}
                \end{center}
                \caption{Heavy CPU Workload Injections}
                \label{fig:impact_contention}
        \end{subfigure}%
        \begin{subfigure}[t]{0.3\textwidth}
                \begin{center}
                        \includegraphics[width=\textwidth]{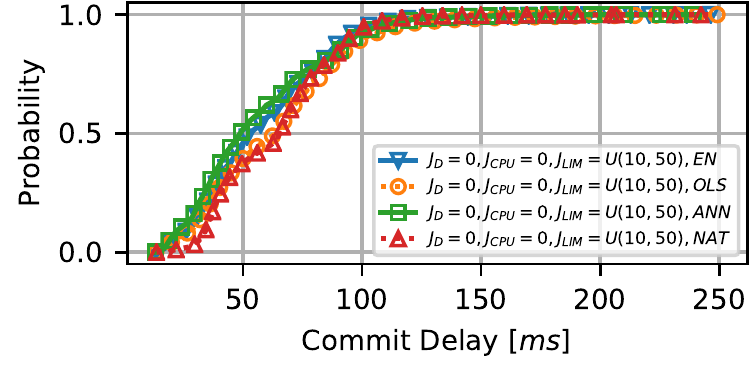}
                        \includegraphics[width=\textwidth]{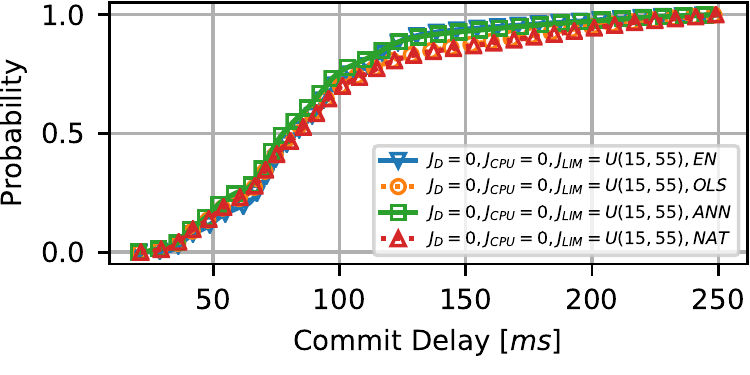}
                        \includegraphics[width=\textwidth]{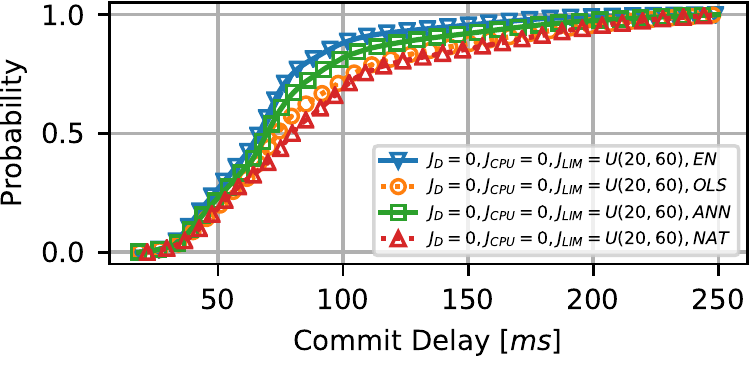}
                \end{center}
                \caption{Heterogeneous CPU Resource Allocations}
                \label{fig:impact_cpu_limit}
        \end{subfigure}
        \caption{ECDFs of observed commit delay for different predictor parametrizations and resource utilization triggers. EN and ANN predictors proved most efficient and consistent across all evaluated scenarios. The relation between CPU and response time is non-linear \cite{wang2005utilization, xu2006predictive}, with OLS thus unsurprisingly performing worse than EN/ANN in scenarios (b) and (c). Interestingly however, OLS underfits the training data in the delay injection scenario (a) as well.}
        \label{fig:plots_impact}
\end{figure*}

\subsubsection{Impact of Delay Injections}
To depict the impact of delay injection on the observed commit delay, we inject uniformly distributed symmetric delays in the range $U(5,35) ms$ in randomly selected replica's attachment ports and observe the predictor performance depicted in Fig. \ref{fig:impact_delay}. Equal amount and intensity of injections are applied both for balanced and imbalanced request arrivals. 

The advantage of \texttt{SEER}'s best-leader prediction and election is observed for all evaluated cluster sizes, with EN's second order polynomial regression resulting in a slightly better resulting performance than ANN in the case of $7$ replicas. For delay injection case and $|\mathcal{R}|=7$, we observe absolute performance advantages of $23.4\%, 18.9\%, 11.3\%$ (mean commit delay) over the Raft's native leader selection (NAT), for EN, ANN and OLS, respectively. Note that the depicted results are valid for Internet2 only. Similar trend was observed for Fat-Tree but was not included here due to space and readability constraints.

\subsubsection{Impact of Resource Contentions}
Fig. \ref{fig:impact_contention} depicts the impact of periodic CPU resource contentions in randomized replica instances. The use of EN predictor in leader election results in mean commit delay decrease of $6.25\%$, $17.1\%$ and $32.07\%$ in $3, 5$, and $7$-replica case respectively, given the Table \ref{other_params} arrival parametrizations.

Indeed, the beneficial performance impact of predictor-based leader election approach is more obvious in the case of a $7$-replica cluster than for the $3$-replica cluster. This is due to fact that for our configured CPU loads, the given combination of resource utilization and client load was insufficient to generate a large impact on the commit performance of the $3$-replica cluster. On average, the $3$-replica cluster generates $3 \cdot \frac{1}{50\cdot10^{-3}} = 60 \frac{req}{s}$, while the $7$-replica cluster generates $7 \cdot \frac{1}{65\cdot10^{-3}} = 107.7 \frac{req}{s}$, under a higher average CPU load for $3$-replica cluster (ref. $J_{CPU}$ in Table \ref{params}) which, when coupled with the corresponding CPU contention values, impacts the cluster in the $7$-replica case more. While we could synthetically optimally overload the $3$-replica cluster to present beneficial results for all cases, we instead make a note here that the beneficial performance of predictor is observed only when a significant CPU load is coupled with sufficiently frequent client request arrivals; or is combined with other imbalance sources (e.g., network delays). 

\subsubsection{Impact of Heterogeneous Res. Allocations}
The CPU resource bounds injected using \texttt{cpulimit} are depicted in Table \ref{other_params} (ref. $J_{LIM}$). As depicted in Fig. \ref{fig:impact_cpu_limit}, EN, ANN and OLS all lead to a similar performance benefit compared to native Raft election (NAT), resulting in mean commit time decrease of $23.9\%$, $15.2\%$ and $5.4\%$ for $7$ replicas, respectively.


\subsection{Impact of Predictor Parametrizations}

\subsubsection{Exploration Trade-Offs}

\begin{figure}[htb]
	\centering
			\includegraphics[width=0.4\textwidth]{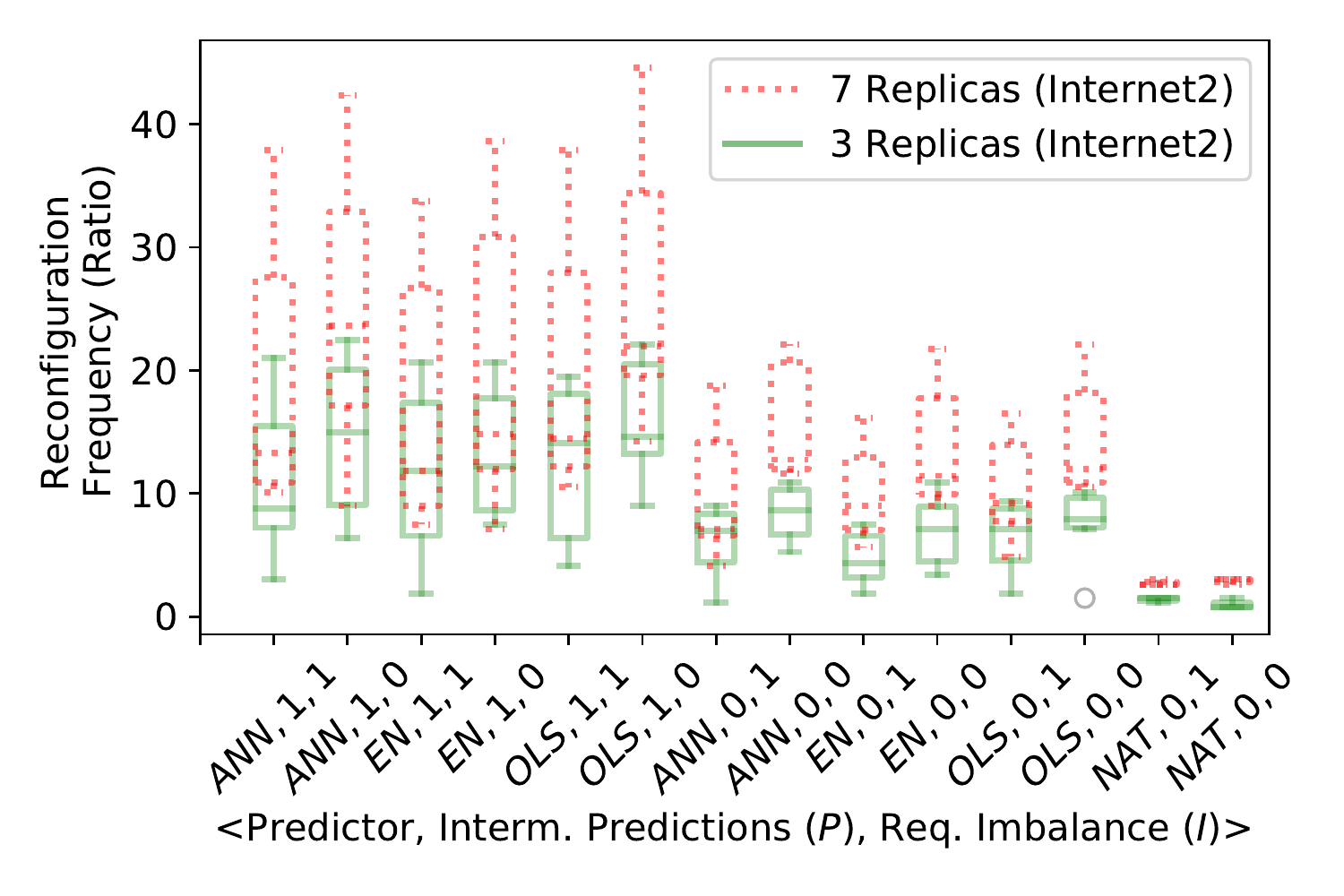}
	\caption{Impact of predictor, intermediate best-leader predictions (LP block), and request imbalance on the leader reconfiguration frequency, normalized to the native Raft leader election case (NAT). Indeed, periodic best-leader predictions (\emph{P=1}) and sparser data collection (\emph{I=1}) result in less total reconfigurations. In return, intermediate predictions benefit the response time (ref. Fig. \ref{fig:prediction_historic_impact}). Predictor choice does not significantly impact the reconf. frequency.}
	\label{fig:plot_reconfigs.pdf}
\end{figure}

In addition to the resource and network dynamics; best-leader prediction and novelty detection intervals $T_P$ and $T_N$, respectively (ref. Fig. \ref{fig:overall_algorithm}) also dictate the expected rate of leader reconfigurations. Re-elections can be associated with brief unavailability periods of servicing user requests. Namely, after announcing a candidate status for an upcoming term, the Raft term is incremented in each instance that receives the candidate request. This results in a transient period during which the requests distributed for confirmation by the leader of the previous term can become rejected and should be resubmitted in the next term. 

Fig. \ref{fig:plot_reconfigs.pdf} depicts the reconfiguration frequency ratio, normalized to the smallest number of reconfiguration measured for the NAT case and $|\mathcal{R}| = 3$. Compared to NAT, \texttt{SEER} comes with the overhead of a higher number of reconfigurations, the actual number linearly scaling with the number of replicas. This is due to a larger number of replicas that must confirm the incremented term prior to leader proposing the client requests. We measure the average leader reconfiguration time in Internet2 to $100.81ms$, $217ms$ and $226.55ms$ for $3$, $5$ and $7$ replicas, respectively. The resulting reconfiguration time in Fat-Tree is slightly lower due to smaller propagation delays associated with the geo-distributed replicas in Internet2.

Compared to the leader election after exploration (i.e., each $T_N$ only), the addition of periodic intermediate predictions each $T_P$ (\emph{P=1}) results in a better observed commit performance. This comes at expense of a higher frequency of updates to the leader's SSM and thus more frequent re-elections (by factor $\sim 2$ for $T_P$, $T_N$, $T_I$ parametrizations as per Table \ref{other_params}).

In contrast to balanced client request arrival distributions (\emph{I=0}), imbalanced distributions (\emph{I=1}) lead to a lower number of leader reconfigurations, which may be associated with a smaller resulting number of matching novel delay commit reports and thus less SSM training samples, leading to less frequent leader elections. \emph{Note:} The number of re-elections for NAT does not equal zero due to potential timeouts of leader messages (i.e., due to leader overload \cite{hanmer2018friend} or network delays). 


\subsubsection{Impact of Frequent Predictions and Rehearsal Learning}

\begin{figure}[htb]
	\centering
	\includegraphics[width=0.48\textwidth]{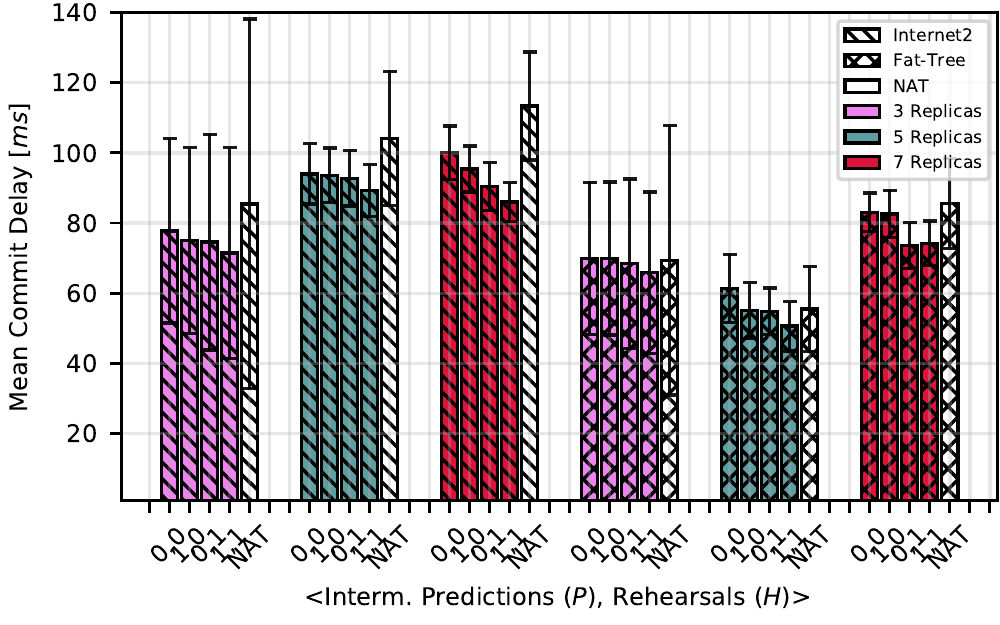}
	\caption{Compared to using intermediate predictions or rehearsal learning individually (\emph{P=1},\emph{H=0} or \emph{P=0,H=1}), enabling both (\emph{P=1,H=1}) consistently results in the lowest response time. Performance increase is observed for both evaluated topologies. The non-NAT bars depict the mean commit delays and corresponding CIs for aggregated EN, OLS, and ANN predictor results.}
	\label{fig:prediction_historic_impact}
\end{figure}

Fig. \ref{fig:prediction_historic_impact} depicts the impact of intermediate Leader Prediction block (LP) execution on the resulting commit delay performance. The LP block elects the best-leader based on current performance predictions for non-follower replicas, without pursuing performance exploration (a task assigned to SSD block). The shown measurements aggregate all parameter combinations, with the uniform Raft leader election mode (NAT) represented by gray bars. The results for ANN, EN, OLS, balanced and imbalanced arrivals are aggregated in individual bars. Compared to Internet2, the reconfig. frequency applied in Fat-Tree is multiplied by $2$, so to allow for overloading the Fat-Tree with remaining parameters unchanged. 

Error bars represent the $\alpha = 0.95$ CIs for the observed commit delay samples. Based on the width of the CIs, we conclude that bottlenecks in smaller clusters generally lead to a larger expected variance of the observed commit delays. A decrease in expected commit delay is observed for \emph{P=1} and \emph{H=1} solutions, an average best-leader result being consistently offered by \emph{P=1,H=1} configuration. \emph{H=1} denotes enabled rehearsal learning during PPM training, where old training sets are uniformly sampled for training vectors. The considered sample quantity ratio for old and new training samples is $1:1$, new samples being the ones collected during actual term. 

We conclude that the application of intermediate prediction leads to better immediate commit performance on average, while the inclusion of old historic data in training becomes beneficial for observed long-term predictor performance. 

\emph{Note}: We have observed no scalability issues during PPM updating, with the ANN model updates generally finishing in $\ll30ms$ on a dedicated x64 CPU core, with $\leq 250$ input observation samples per model.fit() call. Due to paper length limitations we omit including the training overhead results.

\section{Related work}
\label{relatedwork}

	\emph{General Raft protocol optimizations}: Hanmer et al.~\cite{hanmer2019death} demonstrate vulnerability of Raft in a DDoS/overload scenario, leading to frequent leader oscillations. They extend Raft with a "Pre-Vote" round where replicas with expired election timeout first agree on the leader's deteriorated performance prior to electing a new leader to minimize repeated elections. Arora et al.~\cite{arora2017leader} propose relying on quorum-based instead of leader-based reads, so to offload the leader under read-heavy workloads. Kim et al.~\cite{kim2019load} propose an extension to the Raft protocol, where a centralized admission entity approves / rejects per-shard leader elections with the goal of achieving a balanced distribution of leader roles per data shard. Our approach is compatible and orthogonal to these optimizations. 
	
	\emph{Performance optimization via prediction techniques}: Li et al.~\cite{li2015using} propose an ANN-based prediction approach to learn and determine thread prediction strategies for speculative execution of multi-core programs. Shulga et al.~\cite{shulga} propose an ML-based approach to select whether CPU or GPU-based processor should be used for task execution based on input data size. Shekhar et al.~\cite{shekhar} propose an online server selection process that estimates the expected execution time of application tasks on fog and edge nodes and selects those expected to meet the input constraints. The online exploration issue, time-variant replica performance, and distributed nature of resources are ignored in these publications. 
	
	\emph{Leader selection strategies}: Vukolic et al.~\cite{liu2016leader} propose a latency-minimization strategy for imbalanced workloads that reconfigures the leader set of the \emph{all-leader} protocol Clock-RSM~\cite{du2014clock}, based on the previous workload and observed latencies. In~\cite{eischer2018latency}, clients optimize the selection of their leader replicas in a  BFT setting, by sending special probe messages used to collect end-to-end response times. In contrast, we focus on a \emph{single-leader} consensus protocol, additionally consider current resources' utilization as input features, and contrary to these works, pursue a data-driven approach. We are compatible with~\cite{eischer2018latency} in that special probe messages may be generated by the replicas to infer the features used in training and prediction steps when imbalanced per-client arrival distributions are foreseen in the system (e.g., by committing \emph{no-op} log updates). However, we have showcased the effectiveness of \texttt{SEER} for imbalanced loads and have thus omitted the synthetic sample generation. We have also omitted a direct comparison to RTT-only based approaches due to these ignoring the aspects of replica's resource utilization / availability.

\section{Conclusion and Outlook}
\label{conclusion}

We have introduced \texttt{SEER}, an approach to prediction-based leader election in single-leader consensus clusters. \texttt{SEER} was implemented and validated as an extension to a Raft-based data-store in a scenario comprising a replicated routing application, deployed in emulated geo-distributed and local networks using varied predictors and client arrival distributions. In the face of network contentions, contending CPU workloads, and heterogeneous resource allocations, or any combination of the above, compared to uniform leader election, \texttt{SEER} significantly decreases the mean expected cluster response time (by up to $\sim 32\%$ for evaluated scenarios). 

To our best knowledge, this is the first regression-based approach to leader-based election and the first approach to consider non-linear system metrics in the estimation of the best-leader. Most importantly, \texttt{SEER} is deployable without complex modifications to Raft consensus. In fact, it requires minimal modifications to the leader election, in order to enable the external enforcement of candidate roles.

This said, \texttt{SEER} could be improved in two aspects: (i) We currently assume fixed-periodic performance predictions, leader re-elections, and non-leader exploration. An adaptive re-election window could enable additional efficiency and minimize manual parametrization. (ii) We currently observe Raft performance reports for individual, per-application Raft sessions, and accordingly adjust the per-term leader. Future work should exploit common Raft performance metrics across multiple concurrent Raft sessions. Namely, confirming the general effectiveness of \texttt{SEER} without the per-session training observations would possibly benefit its scalability.

\section*{Final Note: Software Artifacts}
Please contact the authors for: (i) the source code of the presented approach including the \texttt{PySyncObj} fork; and (ii) the evaluated data sets and measurement scripts. 


\bibliographystyle{plain}
\bibliography{qos}

\begin{thebibliography}{10}

\bibitem{lofdetection}
{Novelty detection with Local Outlier Factor (LOF)}.
\newblock \url{https://bit.ly/2PabXTf}, 2019.
\newblock [Online; accessed 7-August-2020].

\bibitem{arora2017leader}
Vaibhav Arora et~al.
\newblock {Leader or Majority: Why have one when you can have both? Improving
  Read Scalability in Raft-like consensus protocols}.
\newblock In {\em 9th USENIX Workshop on Hot Topics in Cloud Computing
  (HotCloud 17)}, 2017.

\bibitem{berde2014onos}
Pankaj Berde et~al.
\newblock {ONOS: towards an open, distributed SDN OS}.
\newblock In {\em Proceedings of the third workshop on Hot topics in software
  defined networking}. ACM, 2014.

\bibitem{Breunig:2000:LID:342009.335388}
Markus~M. Breunig et~al.
\newblock {LOF: Identifying Density-based Local Outliers}.
\newblock In {\em Proceedings of the 2000 ACM SIGMOD International Conference
  on Management of Data}, SIGMOD '00. ACM, 2000.

\bibitem{cachin2011introduction}
Christian Cachin et~al.
\newblock {\em Introduction to reliable and secure distributed programming}.
\newblock Springer Science \& Business Media, 2011.

\bibitem{indprofile}
Josef Dorr.
\newblock {IEC/IEEE P60802 JWG TSN Industrial Profile: Use Cases Status Update
  2018-05-14}.
\newblock IEC/IEEE, 2018.

\bibitem{du2014clock}
Jiaqing Du et~al.
\newblock {Clock-RSM: Low-latency inter-datacenter state machine replication
  using loosely synchronized physical clocks}.
\newblock In {\em 2014 44th Annual IEEE/IFIP International Conference on
  Dependable Systems and Networks}. IEEE, 2014.

\bibitem{eischer2018latency}
Michael Eischer et~al.
\newblock {Latency-Aware Leader Selection for Geo-Replicated Byzantine
  Fault-Tolerant Systems}.
\newblock In {\em 2018 48th Annual IEEE/IFIP International Conference on
  Dependable Systems and Networks Workshops (DSN-W)}. IEEE, 2018.

\bibitem{flp}
Michael~J. Fischer et~al.
\newblock {Impossibility of Distributed Consensus with One Faulty Process}.
\newblock {\em Journal of the ACM}, 32(2), April 1985.

\bibitem{ridge}
Gene~H Golub et~al.
\newblock Tikhonov regularization and total least squares.
\newblock {\em SIAM Journal on Matrix Analysis and Applications}, 21(1), 1999.

\bibitem{hanmer2018friend}
Robert Hanmer et~al.
\newblock {Friend or Foe: Strong Consistency vs. Overload in High-Availability
  Distributed Systems and SDN}.
\newblock In {\em 2018 IEEE International Symposium on Software Reliability
  Engineering Workshops (ISSREW)}. IEEE, 2018.

\bibitem{hanmer2019death}
Robert Hanmer et~al.
\newblock {Death by Babble: Security and Fault Tolerance of Distributed
  Consensus in High-Availability Softwarized Networks}.
\newblock In {\em IEEE Conference on Network Softwarization (NetSoft)}. IEEE,
  2019.

\bibitem{hock2014poco}
David Hock et~al.
\newblock {POCO-framework for Pareto-optimal resilient controller placement in
  SDN-based core networks}.
\newblock In {\em 2014 IEEE Network Operations and Management Symposium
  (NOMS)}. IEEE, 2014.

\bibitem{howard2015raft}
Heidi Howard et~al.
\newblock Raft refloated: Do we have consensus?
\newblock {\em SIGOPS Oper. Syst. Rev.}, 49(1), January 2015.

\bibitem{huang2017dynamic}
Xi~Huang et~al.
\newblock {Dynamic switch-controller association and control devolution for SDN
  systems}.
\newblock In {\em 2017 IEEE International Conference on Communications (ICC)}.
  IEEE, 2017.

\bibitem{junqueira2007classic}
Flavio Junqueira et~al.
\newblock {Classic Paxos vs. Fast Paxos: Caveat Emptor}.
\newblock {\em Proceedings of USENIX Hot Topics in System Dependability
  (HotDep)}, 2007.

\bibitem{kemker2018measuring}
Ronald Kemker et~al.
\newblock Measuring catastrophic forgetting in neural networks.
\newblock In {\em Thirty-second AAAI conference on artificial intelligence},
  2018.

\bibitem{kim2019load}
Taehong Kim et~al.
\newblock {Load Balancing of Distributed Datastore in OpenDaylight Controller
  Cluster}.
\newblock {\em IEEE Transactions on Network and Service Management}, 16(1),
  2019.

\bibitem{kingma2014adam}
Diederik~P. Kingma et~al.
\newblock {ADAM: A Method for Stochastic Optimization}.
\newblock In {\em arXiv cs.LG 1412.6980}, 2014.

\bibitem{lamport1998part}
Leslie Lamport.
\newblock The part-time parliament.
\newblock {\em ACM Transactions on Computer Systems (TOCS)}, 16(2), 1998.

\bibitem{li2015using}
Yuxiang Li et~al.
\newblock Using artificial neural network for predicting thread partitioning in
  speculative multithreading.
\newblock In {\em 2015 IEEE 17th International Conference on High Performance
  Computing and Communications, 2015 IEEE 7th International Symposium on
  Cyberspace Safety and Security, and 2015 IEEE 12th International Conference
  on Embedded Software and Systems}. IEEE, 2015.

\bibitem{liu2016leader}
Shengyun Liu et~al.
\newblock Leader set selection for low-latency geo-replicated state machine.
\newblock {\em IEEE Transactions on Parallel and Distributed Systems}, 28(7),
  2016.

\bibitem{Mao:2008:MBE:1855741.1855767}
Yanhua Mao et~al.
\newblock {Mencius: Building Efficient Replicated State Machines for WANs}.
\newblock In {\em Proceedings of the 8th USENIX Conference on Operating Systems
  Design and Implementation}, OSDI'08. USENIX Association, 2008.

\bibitem{odl}
Jan Medved et~al.
\newblock {OpenDaylight: Towards a model-driven SDN controller architecture}.
\newblock In {\em Proceeding of IEEE International Symposium on a World of
  Wireless, Mobile and Multimedia Networks 2014}. IEEE, 2014.

\bibitem{mills1991internet}
David~L Mills.
\newblock Internet time synchronization: the network time protocol.
\newblock {\em IEEE Transactions on communications}, 39(10), 1991.

\bibitem{muja2014scalable}
Marius Muja et~al.
\newblock Scalable nearest neighbor algorithms for high dimensional data.
\newblock {\em IEEE transactions on pattern analysis and machine intelligence},
  36(11), 2014.

\bibitem{nemirovsky2017machine}
Daniel Nemirovsky et~al.
\newblock {A machine learning approach for performance prediction and
  scheduling on heterogeneous CPUs}.
\newblock In {\em 2017 29th International Symposium on Computer Architecture
  and High Performance Computing (SBAC-PAD)}. IEEE, 2017.

\bibitem{netto2017state}
Hylson Netto et~al.
\newblock {State machine replication in containers managed by Kubernetes}.
\newblock {\em Journal of Systems Architecture}, 73, 2017.

\bibitem{ongaro2014search}
Diego Ongaro et~al.
\newblock In search of an understandable consensus algorithm.
\newblock In {\em 2014 USENIX Annual Technical Conference (USENIX ATC 14)},
  2014.

\bibitem{usecases}
{P60802 Project: TSN Profile for Industrial Automation (TSN-IA)}.
\newblock {Use Cases IEC/IEEE 60802 v1.3}.
\newblock IEC/IEEE, 2018.

\bibitem{pedersen2018analysis}
Sebastian Pedersen et~al.
\newblock {An Analysis of Quorum-based Abstractions: A Case Study using Gorums
  to Implement Raft}.
\newblock In {\em Proceedings of the 2018 Workshop on Advanced Tools,
  Programming Languages, and PLatforms for Implementing and Evaluating
  Algorithms for Distributed systems}. ACM, 2018.

\bibitem{pedregosa2011scikit}
Fabian Pedregosa et~al.
\newblock {Scikit-learn: Machine learning in Python}.
\newblock {\em Journal of machine learning research}, 12, 2011.

\bibitem{sakic2018morph}
Ermin Sakic et~al.
\newblock {MORPH: An adaptive framework for efficient and Byzantine
  fault-tolerant SDN control plane}.
\newblock {\em IEEE Journal on Selected Areas in Communications}, 36(10), 2018.

\bibitem{sakicresponse}
Ermin Sakic et~al.
\newblock {Response Time and Availability Study of RAFT Consensus in
  Distributed SDN Control Plane}.
\newblock {\em IEEE Transactions on Network and Service Management}, 15(1),
  March 2018.

\bibitem{shekhar}
S.~{Shekhar} et~al.
\newblock {URMILA: A Performance and Mobility-Aware Fog/Edge Resource
  Management Middleware}.
\newblock In {\em 2019 IEEE 22nd International Symposium on Real-Time
  Distributed Computing (ISORC)}, 2019.

\bibitem{shulga}
D.~A. {Shulga} et~al.
\newblock {The scheduling based on machine learning for heterogeneous CPU/GPU
  systems}.
\newblock In {\em 2016 IEEE NW Russia Young Researchers in Electrical and
  Electronic Engineering Conference (EIConRusNW)}, 2016.

\bibitem{singla2016fat}
Ankit Singla.
\newblock Fat-free topologies.
\newblock In {\em Proceedings of the 15th ACM Workshop on Hot Topics in
  Networks}. ACM, 2016.

\bibitem{lasso}
Robert Tibshirani.
\newblock Regression shrinkage and selection via the lasso.
\newblock {\em Journal of the Royal Statistical Society: Series B
  (Methodological)}, 58(1), 1996.

\bibitem{turcu2014general}
Alexandru Turcu et~al.
\newblock Be general and don’t give up consistency in geo-replicated
  transactional systems.
\newblock In {\em International Conference on Principles of Distributed
  Systems}. Springer, 2014.

\bibitem{veronese2009spin}
Giuliana~Santos Veronese et~al.
\newblock {Spin one's wheels? Byzantine fault tolerance with a spinning
  primary}.
\newblock In {\em 2009 28th IEEE International Symposium on Reliable
  Distributed Systems}. IEEE, 2009.

\bibitem{Vohra2017}
Deepak Vohra.
\newblock {\em {Using Docker in Swarm Mode}}.
\newblock Apress, Berkeley, CA, 2017.

\bibitem{wang2005utilization}
Zhikui Wang et~al.
\newblock Utilization and slo-based control for dynamic sizing of resource
  partitions.
\newblock In {\em International Workshop on Distributed Systems: Operations and
  Management}. Springer, 2005.

\bibitem{linear}
Doug Woos et~al.
\newblock Planning for change in a formal verification of the raft consensus
  protocol.
\newblock In {\em Proceedings of the 5th ACM SIGPLAN Conference on Certified
  Programs and Proofs}, CPP 2016, New York, NY, USA, 2016. Association for
  Computing Machinery.

\bibitem{xu2006predictive}
Wei Xu et~al.
\newblock Predictive control for dynamic resource allocation in enterprise data
  centers.
\newblock In {\em 2006 IEEE/IFIP Network Operations and Management Symposium
  NOMS 2006}. IEEE, 2006.

\bibitem{xu2019elastic}
Zichen Xu et~al.
\newblock {Elastic, Geo-distributed RAFT}.
\newblock In {\em Proceedings of the International Symposium on Quality of
  Service}. ACM, 2019.

\bibitem{zou2005regularization}
Hui Zou et~al.
\newblock {Regularization and variable selection via the Elastic Net}.
\newblock {\em Journal of the royal statistical society: series B (statistical
  methodology)}, 67(2), 2005.

\end{thebibliography}

\end{document}